%% file: main.tex
\documentclass[twocolumn]{aastex631}
\usepackage{printlen}
\input{metrics}

\begin{document}

\title{Improving Generalization and Uncertainty Quantification of Photometric Redshift Models}

\author[0009-0008-4246-0183]{Jonathan Soriano}
\affiliation{Department of Physics and Astronomy, UCLA, Los Angeles, CA 90095, USA}
\author[0000-0001-9554-6062]{Tuan Do}
\affiliation{Department of Physics and Astronomy, UCLA, Los Angeles, CA 90095, USA}
\author[0009-0002-5638-0914]{Srinath Saikrishnan}
\affiliation{Department of Physics and Astronomy, UCLA, Los Angeles, CA 90095, USA}
\author[0009-0006-5531-2126]{Vikram Seenivasan}
\affiliation{Department of Physics and Astronomy, UCLA, Los Angeles, CA 90095, USA}
\author[0000-0001-6790-7297]{Bernie Boscoe}
\affiliation{Computer Science Department, Southern Oregon University, Ashland, OR 97520, USA}
\author[0000-0001-5436-8503]{Jack Singal}
\affiliation{Department of Physics, University of Richmond, Richmond, VA 23173, USA}
\author[0000-0001-7725-2546]{Evan Jones}
\affiliation{Department of Physics and Astronomy, UCLA, Los Angeles, CA 90095, USA}

\begin{abstract}

Accurate redshift estimates are a vital component in understanding galaxy evolution and precision cosmology. In this paper, we explore approaches to increase the applicability of machine learning models for photometric redshift estimation on a broader range of galaxy types. Typical models are trained with ground-truth redshifts from spectroscopy. We test the utility and effectiveness of two approaches for combining spectroscopic redshifts and redshifts derived from multiband ($\sim$35 filters) photometry, which sample different types of galaxies compared to spectroscopic surveys. The two approaches are (1) training on a composite dataset and (2) transfer learning from one dataset to another. We compile photometric redshifts from the COSMOS2020 catalog (TransferZ) to complement an established spectroscopic redshift dataset (GalaxiesML). We used two architectures, deterministic neural networks (NN) and Bayesian neural networks (BNN), to examine and evaluate their performance with respect to the Legacy Survey of Space and Time (LSST) photo-\textit{z} science requirements. We also use split conformal prediction for calibrating uncertainty estimates and producing prediction intervals for the BNN and NN, respectively. We find that a NN trained on a composite dataset predicts photo-\textit{z}'s that are 4.5 times less biased within the redshift range $0.3<z<1.5$, 1.1 times less scattered, and has a 1.4 times lower outlier rate than a model trained on only spectroscopic ground truths. We also find that BNNs produce reliable uncertainty estimates, but are sensitive to the different ground truths. This investigation leverages different sources of ground truths to develop models that can accurately predict photo-\textit{z}'s for a broader population of galaxies crucial for surveys such as Euclid and LSST.         

\end{abstract}

\keywords{Astronomy data analysis (1858) --- Astronomical methods (1043) --- Galaxy Distances (590) --- Galaxy redshifts (631) --- Neural Network (1933)}

\section{Introduction} \label{sec:intro}

In the era of "big data" for astronomy, accurate photometric redshifts (photo-\textit{z}'s) have become essential for measuring the distribution of galaxies across cosmic time, which is essential to many areas of cosmology \citep{salvato2019}. Stage IV galaxy surveys, current and upcoming, such as the ESA's Euclid mission \citep{laureijs2011}, the Vera C. Rubin Observatory's LSST \citep{collaboration2009,ivezic2019}, and the Nancy Grace Roman Space Telescope \citep{akeson2019} depend on reliable estimates of galaxy redshifts from photometry to further our understanding of dark energy and dark matter \citep{mandelbaum2018}. For example, the LSST photo-\textit{z} science goals\footnote{\url{https://docushare.lsstcorp.org/docushare/dsweb/Get/LPM-17}} require (1) a median photo-\textit{z} bias $|\Delta z| < 0.003$, where $\Delta z = (z_{\text{pred}}-z_{\text{true}})/(1+z_{\text{true}})$, (2) a scaled photo-\textit{z} scatter $\sigma_{\Delta z} < 0.02$ \citep{collaboration2009}, and (3) a $3\sigma$ outlier rate below $10\%$ \citep{huterer2006, ma2006, amara2007, kitching2008}. These requirements must be met over a redshift range of $0<z<3$ \citep{collaboration2009}. The Euclid mission requirements are similar to LSST's except over a redshift range $0.2<z<2.5$ and with a stricter median photo-\textit{z} bias $|\Delta z|< 0.002$ \citep{beck2017, collaboration2024}.

Astronomers are continuously improving photo-\textit{z} estimation methods in an attempt to meet these requirements. Given that major surveys such as the Hyper Suprime-Cam Subaru Strategic Program \citep[HSC-SSP;][]{aihara2018} and LSST are designed with only five and six broad bands, respectively, improving photo-\textit{z} estimates derived from few broadband filters is particularly crucial. Photometric redshift methods generally fall into two categories: template-fitting approaches (e.g. \texttt{CPZ}, \citealt{benitez2000}; \texttt{ZEBRA}, \citealt{feldmann2006a}; \texttt{EAZY}, \citealt{brammer2008}; \texttt{LePhare}, \citealt{arnouts2011}) and data-driven approaches (e.g., \texttt{ANNz}, \citealt{collister2004}; \texttt{tpz}, \citealt{carrascokind2013}; \texttt{spiderZ}, \citealt{jones2017}; \texttt{ANNz2}, \citealt{sadeh2016}; \citealt{bonnett2015}; \citealt{jones2024}). These models use measured photometry from astronomical images to produce photo-\textit{z} estimates. Recent approaches predict photo-\textit{z}'s directly from astronomical images \citep{pasquet2019, schuldt2021, jones2024a}. Data-driven methods using machine learning have been shown to meet photo-\textit{z} requirements \citep[e.g.,][]{schmidt2020, newman2022}. A major limitation of data-driven approaches for photo-\textit{z} estimation is the need for training data that are representative of the evaluation data. Typically, training samples use redshift labels derived from spectroscopy. While spectroscopic redshifts provide estimates with uncertainties as low as $\sim10^{-4}$, spectroscopic samples are not representative of the true distribution of galaxies \citep{newman2022}, with bright and emission-line galaxies \citep{beck2017, khostovan2025} being overrepresented. This gap in representativeness between training and target distributions poses a significant challenge for machine learning models.

Photometric redshift catalogs derived from extensive multiwavelength observations can supplement ground truths derived from spectroscopy. These estimates are less precise, but cover a wider range of galaxy types and extend to fainter magnitudes. In this paper, we explore two approaches for incorporating both multiband derived photo-\textit{z}'s and spectroscopic ground truths from real data when training machine learning models: (1) training on a composite dataset and (2) transfer learning \citep{pan2010}. Transfer learning enables models initially trained on broader, less precise datasets to be fine-tuned on precise but narrower data. Transfer learning has only been explored in the past for photometric redshift estimation using simulation to real data \citep{eriksen2020, tanigawa2024}. To the best of our knowledge, few people have used transfer learning from photometric to spectroscopic labels on real data. The composite dataset approach offers an alternative strategy that combines different sources of redshift measurements at the start of training, allowing the model to simultaneously learn from the complementary strengths of spectroscopic and photometric redshifts. Training data that combine spectroscopic data with photometric data have been explored with great success for redshift estimation \citep{tanaka2018,zhou2021,cabayol2023}. We compare these two approaches against conventional models that are trained on a single ground truth.

This work is a continuation of our previous research presented in \citet{soriano2024}, where we developed neural networks (NNs) for photometric redshift estimation using a composite dataset and transfer learning. Here, we extend that framework by evaluating additional photo-\textit{z} metrics and by implementing Bayesian neural networks (BNNs) as these models provide the benefit of uncertainty quantification for predicted photo-\textit{z} estimates. BNNs have been shown to identify outliers in photo-\textit{z} predictions \citep{jones2024a}. This is important in scientific applications, particularly weak-lensing studies, where mischaracterized uncertainties can significantly bias cosmological parameter constraints \citep{huterer2006, hildebrandt2010, hildebrandt2017}. To further improve uncertainty quantification beyond BNNs, we also implement conformal prediction \citep{papadopoulos2011, vovk2012, lei2014, romano2019, vovk2025}, a distribution-free framework that provides mathematically guaranteed coverage. In addition, conformal prediction is capable of producing prediction intervals for nonprobabilistic NNs. Conformal prediction has not been fully explored for photometric redshift regression \citep{jones2024a}.

We have three goals in this work: (1) to develop neural network models that can perform well on a broad population of galaxies, (2) to assess the photo-\textit{z} estimates in the context of LSST, and (3) to investigate the uncertainty quantification using split conformal prediction across two different ground-truth datasets. The paper is structured as follows. In Section \ref{sec:data} we present the data used throughout our analysis. Section \ref{sec:methods} introduces the neural network architectures developed in this work and metrics used to evaluate the two approaches of combining ground truths. In Section \ref{sec:results} we present the performance of the photo-\textit{z} estimates.

\section{Data} \label{sec:data}

\begin{deluxetable*}{lcccccc}
\label{tab:dataset_summary}
\tablecaption{Dataset Summary}
\tablewidth{0pt}
\tabletypesize{\small}
\tablehead{
\colhead{Dataset} & \colhead{$N_{\text{gal}}$} & \colhead{Median $z$} & \colhead{Median $z$ Error} & \colhead{Median \textit{i}-band (mag)} & \colhead{$N_{\text{filt}}$} & MinMax Scale Factor (\textit{g}, \textit{r}, \textit{i}, \textit{z}, \textit{y})
}
\startdata
TransferZ & 116,335 & 0.94 & 0.033 & 23.9 & 5 & 0.065, 0.069, 0.053, 0.055, 0.060 \\
GalaxiesML & 286,401 & 0.49 & 0.00023 & 19.5 & 5 & 0.063, 0.062, 0.069, 0.071, 0.049 \\
Combo & 402,433 & 0.57 & 0.0006 &  20.5 & 5 & 0.050, 0.053, 0.044, 0.045, 0.048
\enddata
\end{deluxetable*}

We utilize two complimentary datasets, TransferZ and GalaxiesML \citep{do2024}, for developing machine learning models for photometric redshift estimation. We created TransferZ from a combination of COSMOS2020 \citep{weaver2022} and the Hyper Suprime-Cam Subaru Strategic Program Second Data Public Release (HSC-PDR2; \citealt{aihara2019}). TransferZ is optimized for machine learning applications (see Section \ref{sec:transferZ}). GalaxiesML was previously created from the HSC-PDR2 Wide Survey matched to spectroscopic redshift surveys. The HSC-SSP team positionally matched ($d\,<\,1^{\prime\prime}$) spectroscopic objects and photometric objects in zCOSMOS \citep{lilly2009}, UDSz \citep{bradshaw2013,mclure2013}, 3D-HST \citep{skelton2014, momcheva2016}, FMOS-COSMOS \citep{silverman2015}, VVDS \citep{fevre2013}, VIPERS PDR1 \citep{garilli2014}, the Sloan Digital Sky Survey (SDSS) DR12 \citep{alam2015}, SDSS DR14Q \citep{paris2018}, GAMA DR2 \citep{liske2015}, WiggleZ DR1 \citep{drinkwater2010}, DEEP2 DR4 \citep{davis2003,newman2013}, DEEP3 \citep{cooper2011,cooper2012}, and PRIMUS \citep{coil2011, cool2013}. For a summary of TransferZ and GalaxiesML, see Table \ref{tab:dataset_summary}.

\subsection{TransferZ}\label{sec:transferZ}
In this work, we create TransferZ, a benchmark photometric redshift dataset. This dataset enables the evaluation of transfer-learning effectiveness and other approaches that combine spectroscopic and photometric redshift ground truths. TransferZ combines 116,335 galaxies with reliable photo-\textit{z}'s from COSMOS2020 with \textit{grizy} photometry (\textit{g}: 4754 \AA, \textit{r}: 6175 \AA, \textit{i}: 7711 \AA, \textit{z}: 8898 \AA, \textit{y}: 9762 \AA) from HSC-PDR2. 

The photometric redshifts in TransferZ are adopted from the COSMOS2020 \texttt{CLASSIC} catalog, derived using the template-fitting code \texttt{LePhare}. The COSMOS2020 catalog consists of two photometric data versions, \texttt{CLASSIC} and \texttt{THE FARMER}, with aperture-based and profile-fitting modeled photometry, respectively. The photo-\textit{z}'s were estimated by two template-fitting algorithms, \texttt{LePhare} \citep{arnouts2011} and \texttt{EAZY} \citep{brammer2008}, modeling spectral energy distributions of 35 broad, intermediate, and narrow bands. The multiwavelength photometry spans the near-ultraviolet to mid-infrared. The COSMOS2020 team derive their HSC \textit{grizy} band photometry from the PDR2 UltraDeep/Deep images. Internally, we find \texttt{CLASSIC}'s \texttt{LePhare} measurements against an HSC-PDR2 spectroscopic sample were marginally more precise and less biased for galaxies with $i<25$ and $0<z<4$. 

Beyond our analysis, the COSMOS2020 team validated their photometric redshifts in each catalog against a large spectroscopic sample \citep{khostovan2025}. The photometric redshifts reach a scatter better than $\sigma_z= 0.025(1+z)$ and $\sigma_z\approx0.05(1+z)$, measured by the normalized mean absolute deviation, for $i_{\text{AB}}<25.0$ and $i_{\text{AB}}<27.0$, respectively. This scatter is worse than with spectroscopic redshift (see Section \ref{sec:ground_truth_compare}), but it is sufficient for many cosmological applications \citep{schmidt2020}. In addition, the COSMOS2020 team find outlier rates, defined by $|\Delta z > 0.15 (1+z_{\text{spec}})$, are lower than 10\% and around 25\% for $i_{\text{AB}}<25.0$ and $i_{\text{AB}}<27.0$, respectively.

The HSC-PDR2 Wide Survey photometry reaches similar depths ($r<26$) as Stage IV surveys like LSST, making it ideal for training and testing machine learning models in preparation for LSST data. HSC-PDR2 covers a smaller area ($\sim$1100 deg$^2$) and observes in one fewer photometric band than LSST, but currently provides the largest sample of galaxy observations comparable to LSST conditions. We note that we use HSC-PDR2 photometry rather than the HSC photometry measurements derived in the COSMOS2020 catalog. This choice ensures that both TransferZ and GalaxiesML use photometric data from the same survey, which is important for consistency in training machine learning models. 

\begin{figure}
    \centering
    \includegraphics[width=0.8\columnwidth]{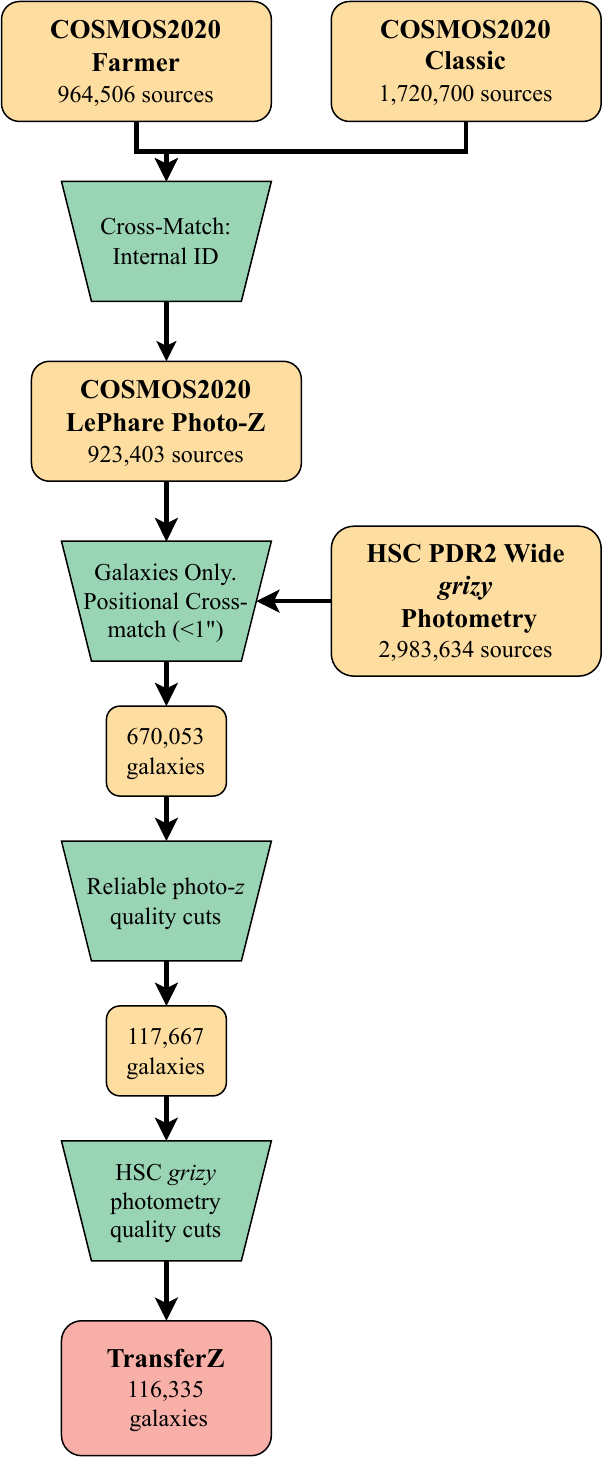}
    \caption{Flowchart showing the steps taken to create TransferZ dataset. Rectangles represent datasets and trapezoids represent processes.}
    \label{fig:transferz_flow}
\end{figure}
\subsubsection{Steps for Building TransferZ}

The first stage in developing TransferZ involved collecting flux data from the HSC-SSP survey and photometric redshifts from COSMOS2020. Initially, we obtained 3 million sources from the HSC-PDR2 Wide survey using HSC-SSP's data-access tools.\footnote{\url{https://hsc-release.mtk.nao.ac.jp/doc/index.php/data-access__pdr3/}} These sources are located within the COSMOS field, centered at R.A. = 10$^{\rm h}$00$^{\rm m}$28$\fs$6 and decl. = +2$^{\circ}$12$^\prime$21$^{\prime\prime}$, (J2000) \citep{scoville2007}. No filtering was applied at this stage. In addition, for $\sim$ 40,000 galaxies we obtained spectroscopic redshifts, which are used for internal validation of COSMOS2020 photometric redshifts \citep[see Appendix A of][]{soriano2024}. We obtained the COSMOS2020 catalog from the publicly available COSMOS team database. 

The TransferZ dataset was created through a filtering process stage designed to ensure quality photometric data. This stage involved four main steps (see Fig. \ref{fig:transferz_flow}):
\begin{enumerate}
    \item We matched sources common to both the Classic and Farmer catalog using a matched identifier provided by the COSMOS2020 team. This resulted in 923,403 sources from COSMOS2020. 
    \item We matched sources common to COSMOS2020 and the HSC-PDR2 sample within 1$^{\prime\prime}$. We also filtered for galaxies using \texttt{LePhare} star--galaxy classification \texttt{lp\_type==0} when available. Otherwise, we used \texttt{i\_extendendness\_value} from HSC-PDR2. In general, both classification methods agree with a statistical precision of $\sim$99\%. This resulted in 670,053 galaxies after filtering. We note the classification filtering we chose removed stars ($\sim$18,000) and active galactic nuclei (AGN; $\sim$3,000) from our sample.
    \item We applied reliable photo-\textit{z} criteria, which resulted in 117,667 galaxies. For more details on the criteria, see Appendix \ref{sec:appendix_a}.
    \item We filtered out bad photometry from HSC-PDR2 across all bands, \texttt{[grizy]\_cmodel\_mag} $<\,50$ and sources with undefined or failed photometry measurements.
\end{enumerate}

After following the steps outlined above, we are left with TransferZ, a dataset of 116,335 galaxies with reliable photometric redshifts in the range $z<4$ and \textit{grizy} photometry from HSC-PDR2. Most sources removed in the second step are fainter than $i=26$, near the detection limit of the PDR2 Wide Survey ($5\sigma$ depth of $i=26.2$), but would still be detectable in the deeper UltraDeep/Deep survey that COSMOS2020 utilizes. Additionally, we believe the AGN removed in Step 2 do not impact this work. After applying similar reliable photometric redshift cuts, accounting for AGN-specific template results in \texttt{LePhare}, the AGN sample is reduced to $\sim350$ AGN.

\begin{figure}[!htbp]
    \centering
    \includegraphics[width=\columnwidth]{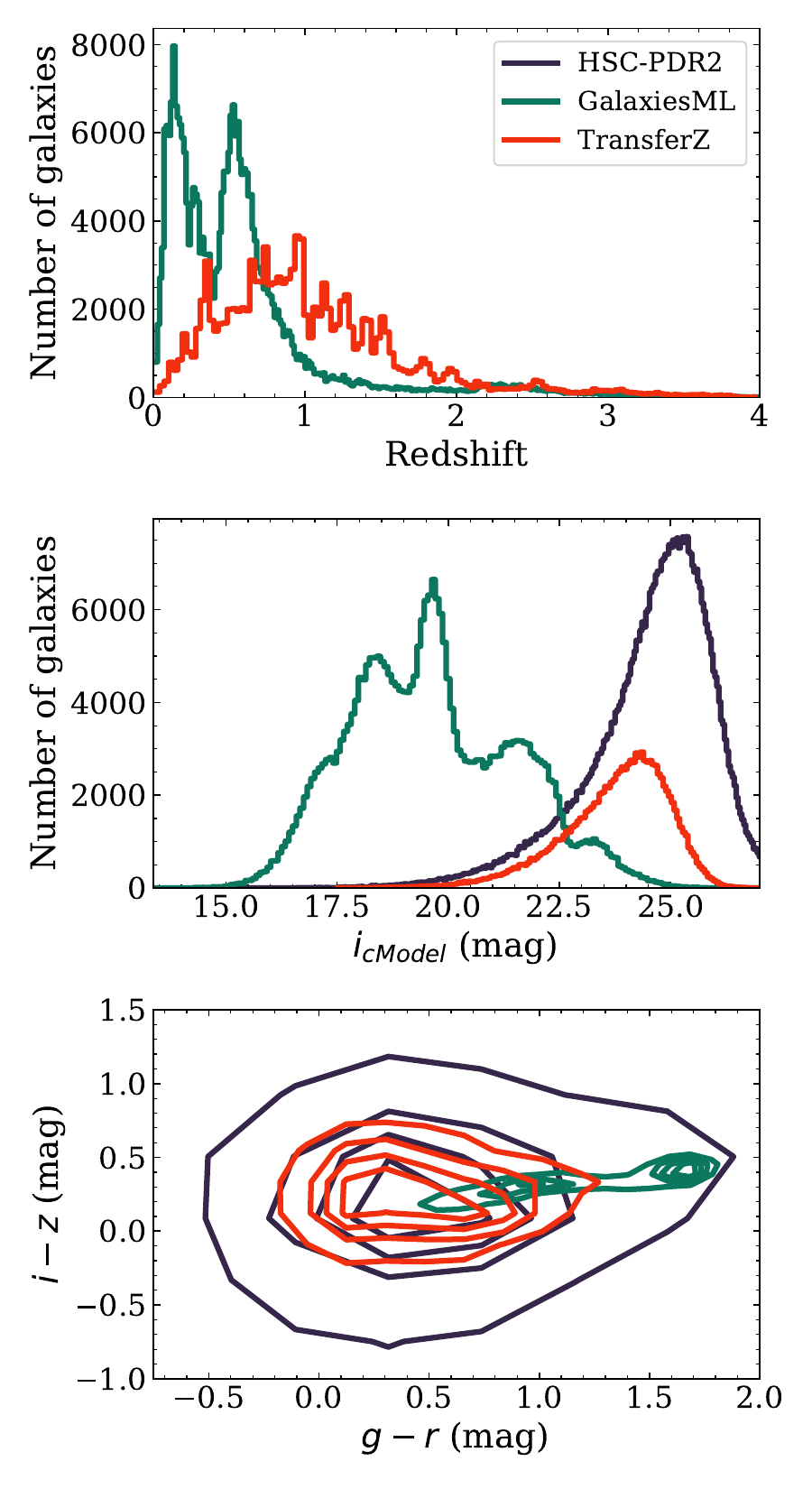}
    \caption{Three datasets: a sample of 500,000 galaxies from HSC-PDR2 
    \citep{aihara2019}, GalaxiesML \citep{do2024} with spectroscopic redshift ground truth, and TransferZ (this work) with COSMOS2020 catalog \citep{weaver2022} multiband imaging redshift ground truth. The distributions of the dataset in redshift (top), \textit{i}-band magnitude (middle), and color-color (bottom) show how the datasets complement each other to help the models generalize beyond the range of brightness and color sampled by a spectroscopic survey.}
    \label{fig:dataset_comparison}
\end{figure}

\subsection{Comparing TransferZ and GalaxiesML}
\label{sec:ground_truth_compare}
TransferZ is a broader and more general galaxy sample that complements GalaxiesML and better represents the HSC-PDR2 parameter space (Fig.~\ref{fig:dataset_comparison}). While GalaxiesML primarily consists of bright galaxies with $i<22$ mag, concentrated at spectroscopic redshifts $z_{\rm spec}<1.2$ with distinct peaks near $z_{\rm spec} \sim 0.1$ and $z_{\rm spec} \sim 0.7$, TransferZ includes fainter galaxies with $i<25$ and photometric redshifts extending to $z_{\rm phot} < 1.9$. The redshift distribution of TransferZ is smoother, spanning a broader range of values. In color-color space, GalaxiesML exhibits a concentrated distribution with multimodal peaks, while TransferZ has a more uniform and wider distribution. TransferZ helps to supplement the parameter space that is underrepresented in GalaxiesML, particularly in the redshift range $0.4 < z < 1.5$ and for fainter galaxies. Although photometric redshifts lack the precision of spectroscopic ones, they allow us to probe a wider, fainter galaxy population. By combining GalaxiesML with TransferZ, we can create a more representative training sample that mitigates the limitations of each dataset individually, avoiding the systematic gaps in parameter space caused by biased sampling.

We investigate methods of combining both sources of ground truth using a composite dataset and transfer learning from one dataset to another. For training on a composite dataset, we combine GalaxiesML and TransferZ into a single dataset, called Combo, of 402,408 galaxies. This approach enables models to learn the full distribution of galaxy properties at once. Approximately 300 galaxies appear in both GalaxiesML and TransferZ, of which we choose to retain the galaxies from GalaxiesML with spec-\textit{z} estimates. In contrast, transfer learning involves pretraining a model on one dataset before fine-tuning it on the other, leveraging knowledge from one problem to adapt to a new problem. Throughout this work, the datasets --- TransferZ, GalaxiesML, and Combo --- are split into 70\% training, 10\% validation, 10\% testing, and 10\% calibration sets. This split differs from machine learning approaches to model development, in which the calibration set serves as calibration for uncertainty quantification via split conformal prediction (see Sec. \ref{sec:conformal}). TransferZ is made available on Zenodo (doi: 10.5281/zenodo.16541823).

\section{Network Architectures and Methodology}\label{sec:methods}

\subsection{Preprocessing}

All models map a five-band magnitude input to a photo-\textit{z} estimate on unseen data. The magnitudes are not extinction corrected as we expect the model to learn this relation. The five-band magnitudes are scaled and translated to a range from 0 to 1 using the minimum and maximum of each band from a model's training set. We present these scaling factors in Table \ref{tab:dataset_summary}.

\subsection{Model Architectures}\label{sec:architectures}

\begin{figure}
    \centering
    \includegraphics[width=0.95\linewidth]{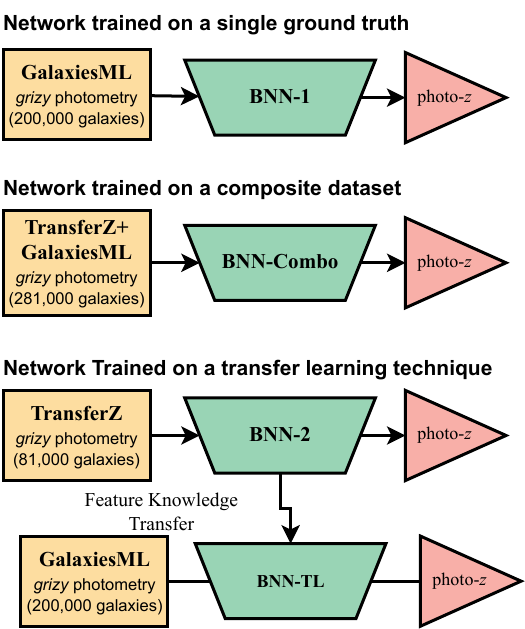}
    \caption{Machine learning training methodologies for photometric redshift estimation using HSC-PDR2 Wide \textit{grizy} photometry. Both deterministic (NN) and probabilistic (BNN) neural networks are trained. Top: baseline networks trained on individual datasets, either GalaxiesML (BNN-1/NN-1) or TransferZ (BNN-2/NN-2). Middle: training on a combination of both datasets (BNN-Combo/NN-Combo). Bottom: transfer-learning methodology with fine-tuning on GalaxiesML from TransferZ pretrained models (BNN-TL/NN-TL).}
    \label{fig:training_tech}
\end{figure}

In our analysis, we used eight NN models: six newly developed models and two reproduced from previous work. These networks include four deterministic NNs, which provide point-estimate redshift predictions, and four BNNs, which output a normal photo-\textit{z} distribution. A normal distribution output may not be appropriate for high-redshift galaxies where multimodal and long-tails probability disitribution functions (pdfs) are common \citep{schmidt2013}. These models were used to test three training conditions: training on a single dataset, training on a composite dataset comprising TransferZ and GalaxiesML, and training using transfer learning (Fig. \ref{fig:training_tech}). We provide visualization of the architectures in Appendix \ref{sec:appendix_b}.

\begin{deluxetable*}{c|cccccc}
\label{tab:nn_search}
\tablecaption{Neural Network Hyperband Search Space and Optimal Values.}
\tablewidth{0pt}
    \tablehead{
    \colhead{Model} & \colhead{No. of} & \colhead{Nodes per} & \colhead{Skip} & \colhead{Additional} & \colhead{No. of} & \colhead{No. of} \\[-2ex]
    \colhead{Type} & \colhead{Hidden Layers} & \colhead{Layer} & \colhead{Connection?} & \colhead{hidden layers?} & \colhead{additional layers} & \colhead{Neurons}
    }
\startdata
    \multicolumn{7}{c}{\textbf{Search Space}} \\
    \hline
    All Models & 1--10 & 32--2048 & True/False & Yes/No & 1--10 & 32--4096 \\
    \hline
    \multicolumn{7}{c}{\textbf{Optimal Values}} \\
    \hline
    NN-1\tablenotemark{a} & 4 & 200, 200, 200, 200 & True & No & -- & -- \\
    NN-2 & 4 & 200, 200, 200, 200 & True & Yes & 2 & 2000, 2000 \\
    NN-Combo & 6 & 1000, 2000, 1000, 2000, 2000, 1000 & True & No & -- & -- \\
\enddata
\tablenotetext{a}{Baseline model reproduced from \citet{jones2024}}
\end{deluxetable*}
\vspace{-24pt}

Our analysis includes neural network models trained on a single ground truth as a baseline for generalization. For a model trained on spectroscopy, we reproduced NN and BNN models from \citet[][hereafter NN-1 and BNN-1, respectively]{jones2024}  that feature four hidden layers with 200 nodes per layer using a rectified linear unit (ReLU) activation with a concatenation-based skipped connection before the output. Both models are trained on the GalaxiesML dataset. To produce a model trained on template-fitting-derived photometric redshifts, we developed NN-2 and BNN-2, which are trained on the TransferZ photometric dataset. These approaches are consistent with most efforts in astronomy to develop machine learning models for photometric redshift estimation.

\begin{deluxetable*}{c|cccc}
\label{tab:bnn_search}
\tablecaption{Bayesian Neural Network Grid Search Space and Optimal Values}
\tablewidth{\textwidth}
    \tablehead{
    \colhead{Model} & \colhead{Architecture\tablenotemark{a}} & \colhead{Prior\tablenotemark{b}} & \colhead{Posterior\tablenotemark{c}} & \colhead{KL Weight\tablenotemark{d}}
    }
\startdata
    \multicolumn{5}{c}{\textbf{Search Space}} \\
    \hline
    All Models & 1--4 & 1--3 & 1--3 & $\{1, 10^{-1}, 10^{-2}, 10^{-3}, 10^{-4}\}$ \\
    \hline
    \multicolumn{5}{c}{\textbf{Optimal Values}} \\
    \hline
    BNN-1\tablenotemark{e} & 1 & 1 & 1 & $10^{-1}$ \\
    BNN-2 & 4 & 1 & 3 & $10^{-2}$ \\
    BNN-Combo & 4 & 1 & 3 & $10^{-2}$ \\
\enddata
\tablenotetext{a}{Architectures: 1 = BNN-1 configuration; 2 = All layers fully variational with BNN-1-like architecture; 3 = Deterministic layers with final variational output layer; 4 = Alternating deterministic and variational layers throughout network}
\tablenotetext{b}{Priors: Gaussian distribution with 1 = trainable $\mu$ and unit $\sigma^2$; 2 = standard norma; or 3 = trainable $\mu$ and $\sigma^2$.}
\tablenotetext{c}{Posteriors: Gaussian distribution with trainable $\mu$ and $\sigma^2$ and 1 = weight initialization; 2 = neural network enhanced; or 3 = neural network enhanced with weight initialization}
\tablenotetext{d}{KL weight values are normalized by the training dataset size, i.e., 0.1/\textit{N}}
\tablenotetext{e}{Baseline model reproduced from \citet{jones2024}}
\end{deluxetable*}
\vspace{-24pt}

We also develop models to test our two approaches for combining ground truths: composite dataset training and transfer learning. The transfer-learning dedicated models, NN-TL and BNN-TL, were fine-tuned to GalaxiesML using NN-2 and BNN-2 as the pretrained models, respectively. We also created NN-Combo and BNN-Combo models, trained on the Combo dataset as our test of composite dataset training.  

For the development of the four unique new models (excluding NN-TL and BNN-TL), we employed hyperparameter tuning strategies for our NNs and BNNs, using NN-1 and BNN-1 architectures as a baseline. For the NNs, we performed a grid search using Hyperband \citep{li2018} to optimize the architectural hyperparameters from NN-1, exploring variations in number of hidden layers, nodes per layer, and concatenation-based skip connections. For a detailed summary, see Table \ref{tab:nn_search}. For the BNNs, we explored both architectural and stochastic model variations. The architectural considerations focused on the implementations of a fully probabilistic BNN to a BNN with deterministic elements. For instance, BNN-1 is a deterministic NN with a single probabilistic layer. The stochastic model variations focus on optimizing the probabilistic components by varying prior distribution, variational posterior distribution, and the Kullback--Leibler (KL) divergence weight. For more details, see Table \ref{tab:bnn_search}.

The hyperparameter optimization yielded unique deterministic NN models, and similar BNN models. For NN-2, as shown in Table \ref{tab:nn_search}, our hyperparameter search identified an optimal architecture that maintains the four hidden layers with 200 nodes per layer from NN-1, but adds two additional layers with 2000 nodes each and implements skip connections. For NN-Combo, we find the optimal architecture features six hidden layers with alternating 1000 and 2000 nodes per layer and skip connections. For the BNN-2 and BNN-Combo, we find similar optimal architectures. Table \ref{tab:bnn_search} shows that both models are designed with a hybrid architecture with alternating deterministic and variational layers throughout the network, using a Gaussian prior with fixed scale parameter, a neural-network-enhanced posterior with controlled initialization strategy, and a KL divergence weight of $10^{-2}$.

\begin{deluxetable*}{l|ccccc}
\label{tab:model_training}
\tablecaption{Neural Network Training Parameters}
\tablewidth{0pt}
    \tablehead{
    \colhead{Model} & \colhead{Training Set} & Training Size & \colhead{Learning Rate} & \colhead{Epochs} & \colhead{Loss Function}
    }
\startdata
    NN-1 & GalaxiesML & 200,000 & $5\times10^{-4}$ & 500 & Eq. \ref{eq:tanka} \\
    BNN-1 & GalaxiesML & 200,000 & $1\times10^{-3}$ & 2000 & NLL \\
    NN-2  & TransferZ & 81,000 & $5\times10^{-4}$ & 500 & Eq. \ref{eq:tanka} \\
    BNN-2 & TransferZ & 81,000 &  $1\times10^{-4}$ & 500 & NLL \\
    NN-Combo & Combo & 281,000 & $1\times10^{-4}$ & 2000 & Eq. \ref{eq:tanka} \\
    BNN-Combo & Combo & 281,000 & $1\times10^{-4}$ & 500 & NLL \\
    NN-TL & GalaxiesML &  200,000 &$5\times10^{-10}$ & 1000 & Eq. \ref{eq:tanka} \\ 
    BNN-TL & GalaxiesML &  200,000 & $5\times10^{-10}$ & 100 & NLL \\
\enddata
\tablecomments{Summary of the training parameters used for each neural network model. NLL refers to the negative log-likelihood loss function. The models with TL suffix used the transfer-learning method with fine-tuning on NN-2 and BNN-2 for the deterministic and probabilistic neural networks, respectively.}
\end{deluxetable*}
\vspace{-24pt}

As aforementioned, two training methodologies are tested in our study. The first uses standard mapping of photometry to redshift estimates using spectroscopic, photometric, or mixed redshift ground truth for all non-transfer-learning models. These were trained using an Adam optimizer \citep{kingma2017} with learning rates on the order of $10^{-4}$. NNs were trained on a custom loss function \citep{tanaka2018}:
\begin{equation}\label{eq:tanka}
    L(\Delta z) = 1-\frac{1}{1+\left(\frac{\Delta z}{\gamma}\right)^2},
\end{equation}
where $\gamma=0.15$ and $\Delta z=(z_{\rm pred}-z_{\rm true})/
(1+z_{\rm true})$. Here, $z_{\rm pred}$ represents predicted point estimates, and $z_{\rm true}$ represents ground-truth redshifts. BNNs were trained on a negative log-likelihood loss. For our BNNs with an \texttt{IndependNormal} output layer, which outputs a Gaussian distribution with predicted mean $\hat{\mu}_z$ and standard deviation $\hat{\sigma}_z$, the negative likelihood loss takes the form

\begin{equation}
    L_{\rm NLL} = \log(\hat{\sigma}_z) + \frac{(z_{\rm true}-\hat{\mu}_{z})^2}{2 \hat{\sigma}_z^2}+\log(\sqrt{2\pi}),  
\end{equation}
where $z_{\rm true}$ is the ground-truth redshift. For completeness, the BNN total training loss is
\begin{equation}
    L_{\rm total} = L_{\rm NLL} + \lambda\times D_{\rm KL}(q_{\phi}||P),
\end{equation}
where $\lambda$ is the KL weight and $D_{\rm KL}$ is the KL divergence of the variational posterior distribution $q_{\phi}$ and the prior $P$ \citep{jospin2022}. This loss function penalizes both prediction  error and uncertainty calibration, enabling the network to learn calibrated uncertainty estimates.

The second methodology implements transfer learning, fine-tuning NN-2 and BNN-2 on GalaxiesML. This approach used a reduced learning rate of $5\times10^{-10}$ for 1000 epochs. For NN-TL, we froze all layers except the input, first, and fifth dense layers, while for BNN-TL only deterministic hidden layers were frozen. A lower learning rate and frozen layers were required to retain information acquired from template fitting photometric redshifts while allowing the model to learn new spectroscopic ground truth. For details on the training parameters, see Table \ref{tab:model_training}.

All models are implemented in TensorFlow/Keras \citep{abadi2016}. For the BNNs, we use TensorFlow Probability \citep{dillon2017} with variational inference to handle posterior distributions. 

\subsection{Photo-\textit{z} Performance Metrics}
In this section, we present the metrics utilized to evaluate our photometric redshift estimates. We first provide a description of our point-estimate metrics (Section \ref{sec:point_metrics}). We then proceed to describe our uncertainty quantification procedure via split conformal prediction in Section
\ref{sec:conformal} followed by our metrics for the uncertainty estimates in Section \ref{sec:prob_metrics}.

\subsubsection{Point-estimate Metrics}\label{sec:point_metrics}
Our choice of metrics is consistent with those required for cosmological studies in large-scale surveys like LSST and Euclid. The LSST Science Requirements document defines photometric redshift requirements for galaxies with $i<25.3$ as (1) bias ($b\,<\,0.003$) (2) scatter ($\sigma_z/(\,1+\,z)<0.02$), and (3) a $3\sigma$ outlier rate below 10\%. Uncertainty estimates produced by BNNs and split conformal prediction (see Sec \ref{sec:conformal}) serve to compute the $3\sigma$ outlier rates. The definitions used for these metrics are consistent with the literature \citep{hildebrandt2012,dahlen2013,tanaka2018,salvato2019,desprez2020,singal2022}

We evaluate the accuracy of the photo-\textit{z} estimates by the redshift weighted difference between the photo-\textit{z} and ground-truth redshifts, defined by
\begin{equation}
    \Delta z=\frac{z_{\rm pred}-z_{\rm true}}{1+z_{\rm true}},
\end{equation}
where $z_{\rm pred}$ is the predicted point estimate or median from the photo-\textit{z} distribution from the NNs and BNNs, respectively, and $z_{\rm true}$ is the ground-truth redshifts. We define the bias as the median of the $\Delta z$ distribution, $b=\text{median}(\Delta z)$, and scatter as 
\begin{equation}
    \sigma_{\rm NMAD}= 1.48\times\text{MAD}|\Delta z|,
\end{equation}
 which is less sensitive to outliers than the alternative definition of the standard deviation of the $\Delta z$ distribution. For sources with uncertainty estimates, the $3\sigma$ outlier rate ($O_{3\sigma}$) is defined as the percentage of sources for which $|z_{\rm pred}-z_{\rm true}| > 3\sigma$. Alternatively, we provide the metric $O$ which is the percentage of sources satisfying $|\Delta z| > 0.15$, where the threshold value is appropriate for our setup of several band-derived photometric redshifts \citep{tanaka2018}. In addition, we provide the catastrophic outlier rate, $CO$, defined by \citet{singal2022} as the percentage of sources for which $|z_{\rm pred}-z_{\rm true}|>1.0$.  

We evaluate the performance of the models on each testing dataset within the range $0.3<z<1.5$, a range relevant to cosmological probes. The metrics are evaluated on average over this range. Weak-lensing and other cosmological analyses require the performance of photo-\textit{z} estimates to be achieved on average throughout tomographic bins. However, an average performance over the range provides a simple comparison between our models. 

\subsubsection{Split conformal prediction}\label{sec:conformal}
Conformal prediction \citep{vovk2012,lei2014} provides a distribution-free, model-agnostic framework for constructing prediction intervals to achieve a guaranteed statistical coverage level (e.g., 68\%). It enhances the probabilistic outputs of BNNs, and predictions by NNs are assigned predictiion intervals \citep{vovk2012}. Assuming our data are \textit{exchangeable} (a weaker condition than i.i.d.), prediction bands for a new observation are formed using by calibrating on the errors from past observations. In particular, this work applies \textit{split conformal prediction}, a variant of conformal prediction that sets aside a fraction of the data for the calibrations. While this reduces the amount of training data, it is computationally efficient.

Since model uncertainties vary with redshift, particularly in high-redshift regions with sparse training data, we apply split conformal prediction within tomographic bins of width $\delta z=0.1$. The ground-truth redshift is used to assign galaxies to their respective band. This assignment can be problematic for new sources without ground-truth labels, but this is beyond the scope of the present paper. Using a calibration set (10\% of each dataset), we calculate nonconformity scores. For BNNs, these scores are normalized by the predicted standard deviation:
\begin{equation}
    E_{i}=\frac{|y_{i}-\hat{\mu}(x_{i})|}{\hat{\sigma}(x_{i})},
\end{equation}
where $y_{i}$ is the ground truth, $\hat{\mu}(x_{i})$ is the predicted mean, and $\hat{\sigma}(x_{i})$ is the predicted standard deviation. For NNs, which only provide point estimates, nonconformity scores are defined as
\begin{equation}
    E_{i}=|y_{i}-\hat{\mu}(x_{i})|.
\end{equation}

We then compute the conformal correction factor, $\hat{q}_{i}$ from the 68\% quantile, $Q$, of these scores for each redshift tomographic bin: $\hat{q}_i=Q(\{E_i\}, 0.68)$. For BNN uncertainty estimates, the resulting calibrated 68\% prediction interval for an individual galaxy belonging to the $i$th tomographic redshift bin is
\begin{equation}
    \tau(x_{j})=[\hat{\mu}(x_{j})-\hat{q}_i\hat{\sigma}(x_{j}), \hat{\mu}(x_{j})+\hat{q}_i\hat{\sigma}(x_{j})].
\end{equation}
This split conformal prediction approach corrects each uncertainty estimate from the BNN by a factor of $\hat{q}_i$ in the $i$th redshift bin. For the NNs, which do not produce uncertainty estimates, the split conformal prediction yields 68\% prediction intervals for individual galaxies belonging to the $i$th tomographic redshift bin as
\begin{equation}
    \tau(x_{j})=[\hat{\mu}(x_{j})-\hat{q}_i, \hat{\mu}(x_{j})+\hat{q}_i].
\end{equation}
The prediction interval width is constant for all galaxies within a given redshift bin, as split conformal prediction quantifies bin-level uncertainty rather than individual galaxy uncertainty estimates 

For completeness, we denote the catastrophic outlier fraction for split conformal prediction calibrated uncertainties, using the $3\sigma$ definition, as $O_{3\sigma,\rm cp}$. We note that the correspondence between the 68\% prediction interval and $1\sigma$ (and the derived $3\sigma$ threshold) assumes the underlying error distribution is approximately Gaussian. For the $O_{3\sigma, \rm cp}$ metric, this should be interpreted as the width of the prediction interval rather than a formal standard deviation of a Gaussian distribution. This approximation is appropriate for practical comparison with LSST science requirements, though the $3\sigma$ outlier rate should be interpreted as an order-of-magnitude evaluation of extreme outliers rather than a formal statistical quantity.

\subsubsection{Uncertainty Evaluation Metrics}\label{sec:prob_metrics}

To evaluate the accuracy and reliability of probabilistic redshift predictions from our BNNs after applying split conformal prediction, we use complementary metrics: statistical coverage and global probability integral transform (PIT) plots. For a comparison before and after split conformal prediction, see Appendix \ref{sec:appendix_c}. 

Statistical coverage quantifies whether prediction intervals accurately reflect their nominal level. Specifically, we define statistical coverage as the proportion of galaxies whose true redshift ($z_{\rm true}$) falls within the predicted 68\% prediction interval:

\begin{equation}
    \text{Coverage} = \frac{1}{N}\sum_{i=1}^{N} |z_{pred,i}-z_{true,i}| < \sigma_{i},
\end{equation}
where $N$ is the total number of galaxies in the testing set. Ideal calibration yields statistical coverage of exactly 68\%. When statistical coverage exceeds 68\%, the prediction intervals are too wide (overcover), while statistical coverage below 68\% indicates intervals that are too narrow (undercover). We evaluate statistical coverage within tomographic bins of size $\Delta z=0.1$ to capture redshift-dependent trends.

\begin{figure*}[!htp]
    \centering
    \includegraphics[width=\textwidth]{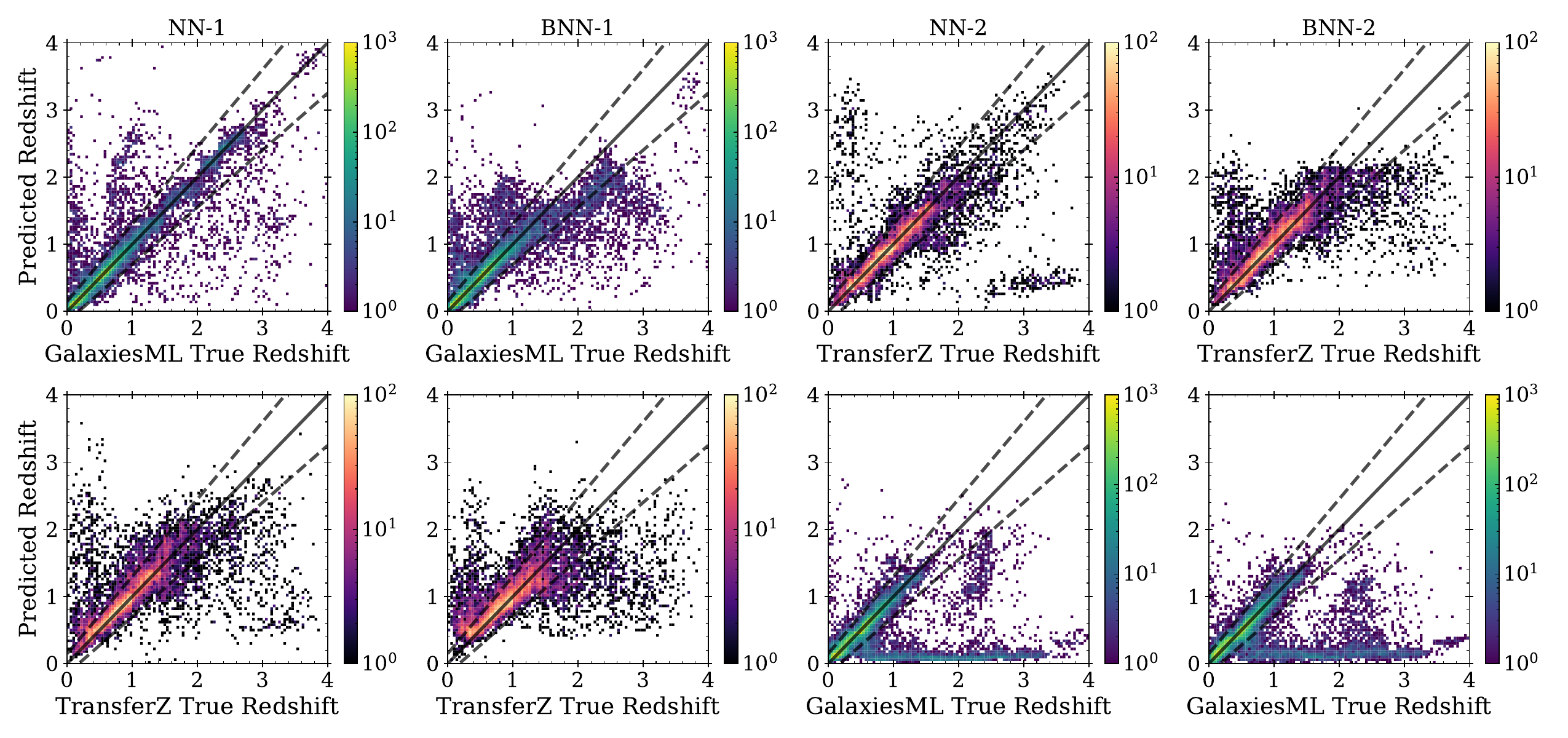}
    \caption{Photometric vs. ground-truth redshift for deterministic and probabilistic models trained and evaluated on different datasets. The top row shows the models evaluated on a sample similar to the training dataset. The bottom row shows the models evaluated on a sample different from the training dataset. NN-1 and BNN-1 were trained on 200,000 galaxies from GalaxiesML. NN-2 and BNN-2 were trained on 81,000 galaxies from TransferZ. The solid black line is a one-to-one line and the dashed black lines correspond to outliers with $z_{\rm pred}>\pm0.15(1\,+\,z_{\rm true})$. The color scale indicates the density of the data points (log scale). All models perform well on the ground truth they are trained on, but show greater scatter and more outliers on a sample different from their training set.}
    \label{fig:single_pred}
\end{figure*}

The PIT provides a visual diagnostic for assessing whether predictive distributions are well calibrated \citep{dey2022a}. The PIT value is the cumulative distribution function evaluated at the true redshift for each source, and is defined as 
\begin{equation}
    \mathrm{PIT}(z_{\rm true})= \int_{0}^{z_{\rm true}} p(z)\,dz
\end{equation}
where $p(z)$ is the predicted pdf of a source \citep{schmidt2020}. The distribution of PIT values probes the performance of photo-\textit{z} and their uncertainty estimates (pdfs) over an ensemble of galaxies. For accurate redshift estimates and perfectly calibrated uncertainty measurements, the PIT distribution for a sample should follow a uniform distribution \textit{U}[0,1].

The morphology of a PIT histogram reveals specific calibration issues:
\begin{itemize}
    \item A U-shaped histogram indicates narrow or underdispersed pdfs (overconfidence).
    \item An inverted U-shape (center-peaked) histogram indicates broad or overdispersed pdfs (underconfidence).
    \item Asymmetric or sloped histograms reveal systematic biases in the predicted distributions.
    \item Concentrated mass at extremes (PIT = 0 or 1) indicates the presence of outliers, where the true redshift falls outside the predicted distribution.
\end{itemize}

Global PIT is known to be an imperfect diagnostic for regression problems with heterogeneous data. For photo-\textit{z} models, the global PIT can be unreliable i.e., it can display a uniform PIT distribution when the calibration is poor \citep{schmidt2020, zhao2021}. Thus, PIT results should not be interpreted as definitive evidence of calibration quality. More robust diagnostics, such as conditional or local PIT evaluated within bins of magnitude or redshift \citep{zhao2021, dey2022c}, may provide a clearer interpretation and are left for future work.

Since split conformal prediction yields only prediction intervals, we compute PIT values by treating the calibrated 68\% intervals as Gaussian-distributed uncertainties. Specifically, for each galaxy with predicted redshift $z_{\rm pred}$ and conformal prediction interval half-width $\sigma_{\rm cp}$, we construct a normal distribution and approximate  the PIT  as 
\begin{equation}
    \mathrm{PIT}(z_{\rm true})=\frac{1}{2}\left[1+\mathrm{erf}\left(\frac{z_{\rm true}-z_{\rm pred}}{\sigma_{\rm cp}\sqrt{2}}\right)\right].
\end{equation}
This approach provides a consistent mapping for qualitative PIT assessment, but has several important limitations. First, it assumes an underlying Gaussian error distribution, which may not reflect the true error structure. Second, it does not capture the full non-Gaussian predictive density that may be present in photometric redshift estimation. Therefore, the PIT results presented here should be interpreted cautiously and not as definitive evidence of calibration quality.

\section{Results} \label{sec:results}

In this section, we characterize the performance of neural networks and different approaches to mixing ground truths for photometric redshift estimation. We begin by evaluating the precision and generalizability of our models using the metrics detailed in Sections \ref{sec:point_metrics} and \ref{sec:prob_metrics}, as well as evaluating the models in the context of LSST science requirements. For this work, we define precision as a single ground-truth trained model's photometric redshift accuracy on data resembling its training sample. Generalizability is defined as a model's ability, trained on a single ground-truth, to accurately predict photometric redshift on data different from its training sample. A model trained on mixed ground truths achieves generalizability if its performance remains consistent across different ground-truth types. In addition to point-estimate performance, we analyze the effectiveness of uncertainty quantification methods across our models with split conformal prediction.  Throughout our evaluation, we highlight the contrast between deterministic NNs and BNNs. In Appendix \ref{sec:metric_per}, we provide metric results per redshift and magnitude. Additionally, in Appendix  \ref{sec:mrt}, Table \ref{tab:predictions} lists the redshift estimates for 40,269 galaxies in the GalaxiesML and TransferZ datasets using our eight models. 

\begin{deluxetable*}{l|ccccccc}\label{tab:metric_fnn}
\tablecaption{NNs: Averaged Metrics over $0.3<z<1.5$}
\tablewidth{0pt}
    \tablehead{
    \multicolumn{1}{l|}{Model} & \colhead{Training Data\tablenotemark{a}} & \colhead{Testing Data\tablenotemark{b}} & \colhead{$|b|$} & \colhead{$\sigma_{\rm NMAD}$} & \colhead{$O$} & $CO$ & \colhead{$O_{3\sigma,\rm cp}$} \\[-2ex]
    & & & \colhead{($\times10^{-3}$)} & \colhead{($\times10^{-2}$)} & \colhead{($\%$)} & \colhead{($\%$)} & \colhead{($\%$)} 
    }
\startdata
    \multicolumn{8}{c}{\textbf{Precision}} \\
    \hline
    NN-1 & GalaxiesML & GalaxiesML & \nnBiasGMLone & \nnScatGMLone & \nnOGMLone & \nnCOGMLone &  \nnOsigConfGMLone \\
    NN-2 & TransferZ & TransferZ   & \nnBiasTZtwo & \nnScatTZtwo & \nnOTZtwo & \nnCOTZtwo &  \nnOsigConfTZtwo  \\
    \hline
    \multicolumn{8}{c}{\textbf{Generalizability}} \\    
    \hline
    NN-1 & GalaxiesML & TransferZ   & \nnBiasTZone & \nnScatTZone & \nnOTZone & \nnCOTZone &  \nnOsigConfTZone \\
    NN-1 & GalaxiesML & Combo       & \nnBiasCone & \nnScatCone & \nnOCone & \nnCOCone &  \nnOsigConfCone \\
    NN-2 & TransferZ & GalaxiesML   & \nnBiasGMLtwo & \nnScatGMLtwo & \nnOGMLtwo & \nnCOGMLtwo &  \nnOsigConfGMLtwo \\
    NN-2 & TransferZ & Combo        & \nnBiasCtwo & \nnScatCtwo & \nnOCtwo & \nnCOCtwo &  \nnOsigConfCtwo \\
    NN-TL\tablenotemark{c}  & GalaxiesML & GalaxiesML & \nnBiasGMLtl & \nnScatGMLtl & \nnOGMLtl & \nnCOGMLtl &  \nnOsigConfGMLtl \\
    NN-TL & GalaxiesML & TransferZ  & \nnBiasTZtl & \nnScatTZtl & \nnOTZtl & \nnCOTZtl &  \nnOsigConfTZtl \\
    NN-TL & GalaxiesML & Combo      & \nnBiasCtl & \nnScatCtl & \nnOCtl & \nnCOCtl &  \nnOsigConfCtl \\
    NN-Combo  & Combo & Combo        & \nnBiasCcom & \nnScatCcom & \nnOCcom & \nnCOCcom &  \nnOsigConfCcom \\
    NN-Combo & Combo & GalaxiesML   & \nnBiasGMLcom & \nnScatGMLcom & \nnOGMLcom & \nnCOGMLcom &  \nnOsigConfGMLcom \\
    NN-Combo & Combo & TransferZ    & \nnBiasTZcom & \nnScatTZcom & \nnOTZcom & \nnCOTZcom & \nnOsigConfTZcom \\
    \hline
    LSST Requirements & & & $<3.00$ & $<2.0$ & & & $<10.0$ \\
    \hline 
\enddata
\tablenotetext{a}{The training size of each dataset are: GalaxiesML, 200,000; TransferZ, 81,000; Combo, 281,000.}
\tablenotetext{b}{The testing size of each dataset are: GalaxiesML, 28,000; TransferZ, 11,000; Combo, 41,000.}
\tablenotetext{c}{NN-TL is fine-tuned on GalaxiesML, with NN-2 serving as the baseline model.}
\end{deluxetable*}

\begin{deluxetable*}{l|ccccccc}\label{tab:metric_bnn}
\tablecaption{BNNs: Averaged Metrics over $0.3<z<1.5$}
\tablewidth{0pt}
    \tablehead{
    \multicolumn{1}{l|}{Model} & \colhead{Training Data\tablenotemark{a}} & \colhead{Testing Data\tablenotemark{b}} & \colhead{$|b|$} & \colhead{$\sigma_{\rm NMAD}$} & \colhead{$O$} & $CO$ & \colhead{$O_{3\sigma,\rm cp}$} \\[-2ex]
    & & & \colhead{($\times10^{-3}$)} & \colhead{($\times10^{-2}$)} & \colhead{($\%$)} & \colhead{($\%$)} & \colhead{($\%$)}
    }
\startdata
    \multicolumn{8}{c}{\textbf{Precision}} \\
    \hline
    BNN-1 & GalaxiesML & GalaxiesML   & \BnnBiasGMLone & \BnnScatGMLone & \BnnOGMLone & \BnnCOGMLone & \BnnOsigConfGMLone \\
    BNN-2 & TransferZ & TransferZ     & \BnnBiasTZtwo & \BnnScatTZtwo & \BnnOTZtwo & \BnnCOTZtwo & \BnnOsigConfTZtwo \\
    \hline
    \multicolumn{8}{c}{\textbf{Generalizability}} \\    
    \hline
    BNN-1 & GalaxiesML & TransferZ   & \BnnBiasTZone & \BnnScatTZone & \BnnOTZone & \BnnCOTZone & \BnnOsigConfTZone \\
    BNN-1 & GalaxiesML & Combo       & \BnnBiasCone & \BnnScatCone & \BnnOCone & \BnnCOCone & \BnnOsigConfCone \\
    BNN-2 & TransferZ & GalaxiesML   & \BnnBiasGMLtwo & \BnnScatGMLtwo & \BnnOGMLtwo & \BnnCOGMLtwo & \BnnOsigConfGMLtwo \\
    BNN-2 & TransferZ & Combo        & \BnnBiasCtwo & \BnnScatCtwo & \BnnOCtwo & \BnnCOCtwo & \BnnOsigConfCtwo \\
    BNN-TL\tablenotemark{c}  & GalaxiesML & GalaxiesML & \BnnBiasGMLtl & \BnnScatGMLtl & \BnnOGMLtl & \BnnCOGMLtl & \BnnOsigConfGMLtl \\
    BNN-TL & GalaxiesML & TransferZ  & \BnnBiasTZtl & \BnnScatTZtl & \BnnOTZtl & \BnnCOTZtl & \BnnOsigConfTZtl \\
    BNN-TL & GalaxiesML & Combo      & \BnnBiasCtl & \BnnScatCtl & \BnnOCtl & \BnnCOCtl & \BnnOsigConfCtl \\
    BNN-Combo  & Combo & Combo        & \BnnBiasCcom & \BnnScatCcom & \BnnOCcom & \BnnCOCcom & \BnnOsigConfCcom \\
    BNN-Combo & Combo & GalaxiesML   &  \BnnBiasGMLcom & \BnnScatGMLcom & \BnnOGMLcom & \BnnCOGMLcom & \BnnOsigConfGMLcom \\
    BNN-Combo & Combo & TransferZ    & \BnnBiasTZcom & \BnnScatTZcom & \BnnOTZcom & \BnnCOTZcom & \BnnOsigConfTZcom \\
    \hline
    LSST Requirements & & & $<3.00$ & $<2.00 $ & & & $<10.0$ \\
    \hline 
\enddata
\tablenotetext{a}{The training size of each dataset are: GalaxiesML, 200,000; TransferZ, 81,000; Combo, 281,000.}
\tablenotetext{b}{The testing size of each dataset are: GalaxiesML, 28,000; TransferZ, 11,000; Combo, 41,000.}
\tablenotetext{c}{BNN-TL is fine-tuned on GalaxiesML, with BNN-2 serving as the baseline model.}
\end{deluxetable*}

\begin{figure}
    \centering
    \includegraphics[width=\columnwidth]{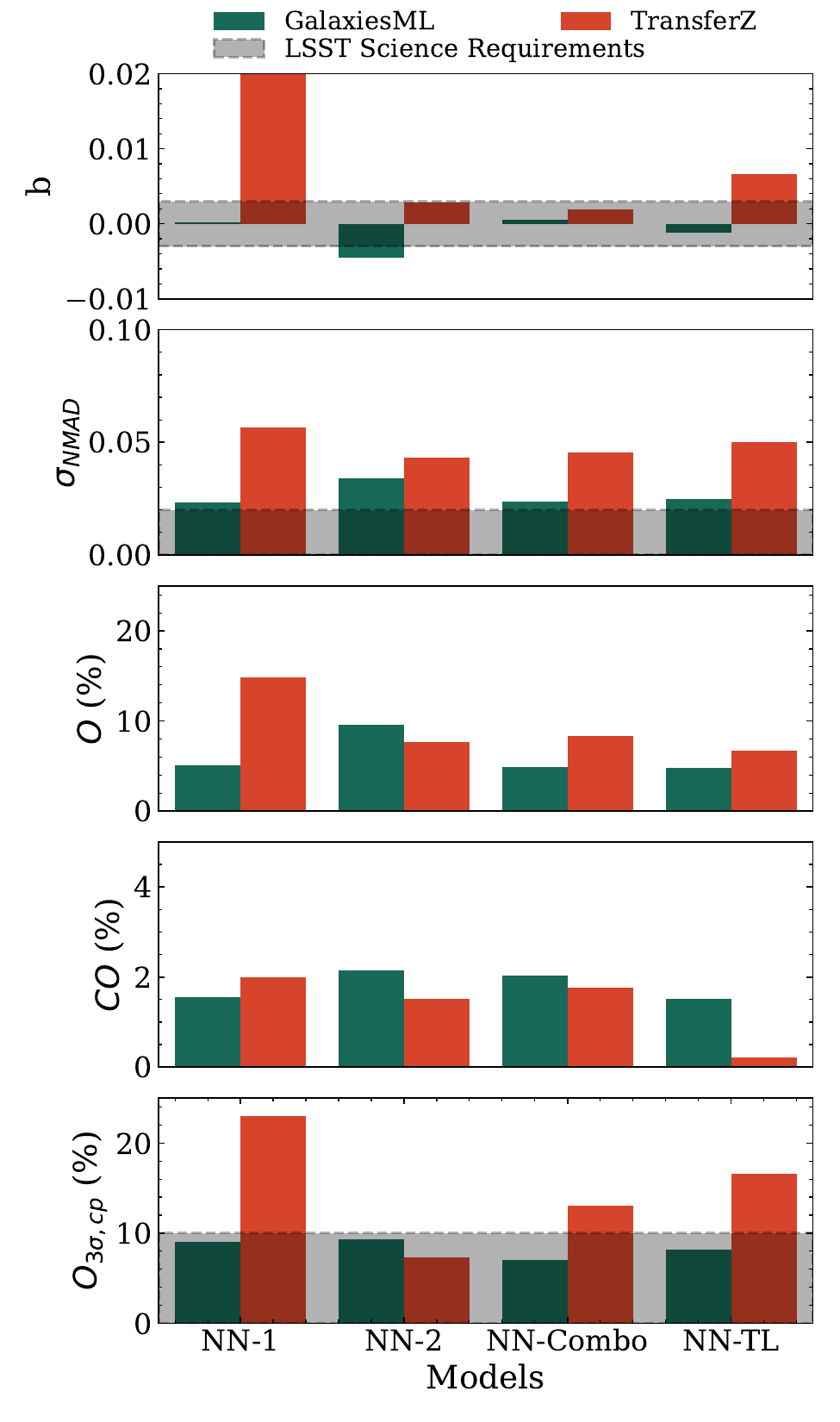}
    \caption{A comparison of the performance of NNs using metrics averaged over the redshift range $0.3<z<1.5$ (see Sec. \ref{sec:point_metrics}). The gray shaded area corresponds to the LSST science goals. The NN-TL and NN-Combo models show comparable performance on both GalaxiesML and TransferZ, with NN-Combo meeting one more requirement than NN-TL. NN-2 performs comparably to NN-TL and NN-Combo, but with a higher outlier rate $O$ on GalaxiesML.}
    \label{fig:nn_metric_comparison}
\end{figure}

\begin{figure}
    \centering
    \includegraphics[width=\columnwidth]{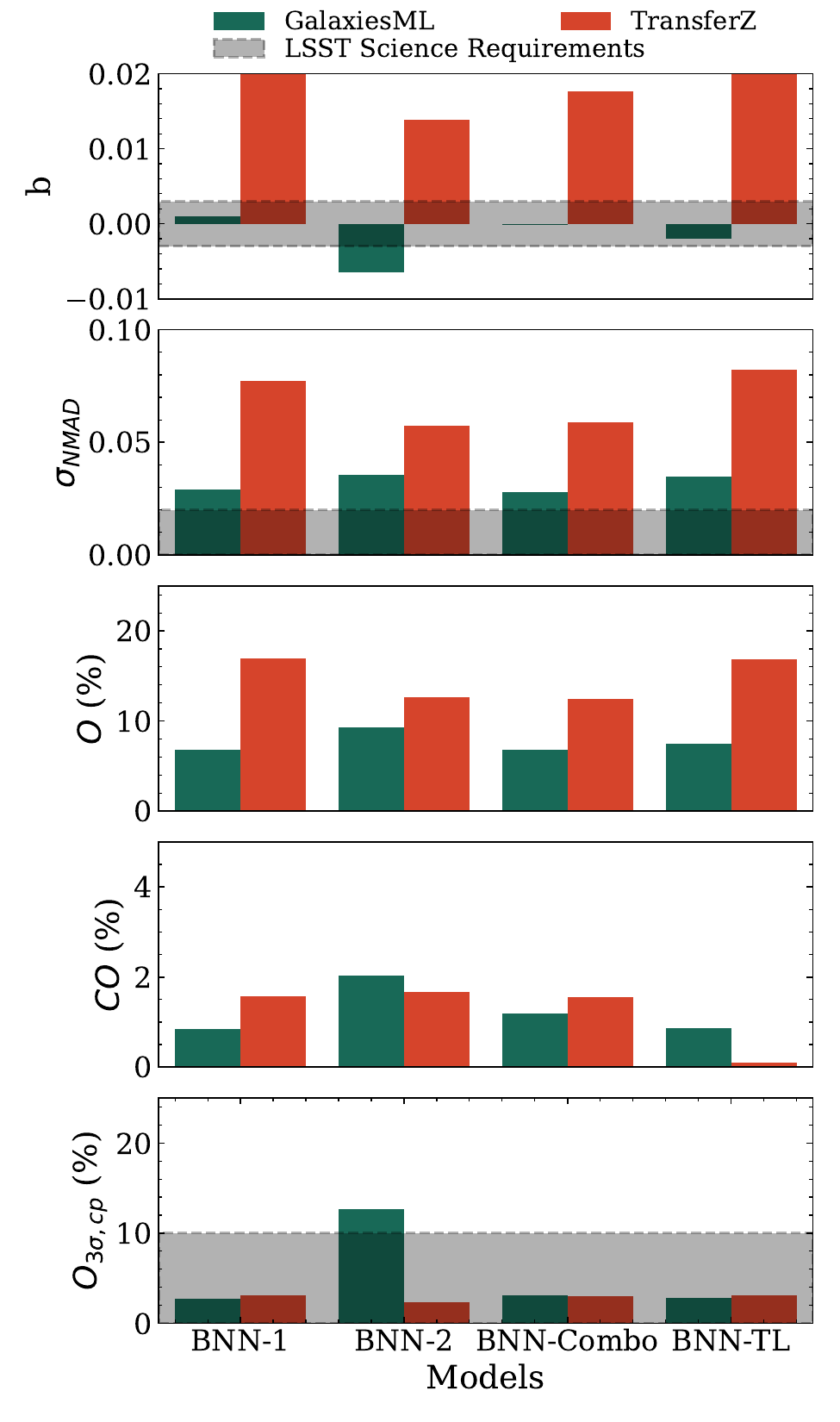}
    \caption{A comparison of the performance of BNNs using metrics averaged over the redshift range $0.3<z<1.5$ (see Sec. \ref{sec:point_metrics}). The gray shaded area corresponds to the LSST science goals. The models, with the exception of BNN-2, meet the same LSST bias and outlier requirements. Regardless of the training approach, these models fail to perform comparably on both ground truths.}
    \label{fig:bnn_metric_comparison}
\end{figure}

\subsection{Photo-\textit{z} Precision: Training and Evaluating on Similar Data}
We report the baseline performance, or precision, of models evaluated on the same ground truth as their training sample. The top row of Figure \ref{fig:single_pred} shows the predictions by NN-1 and BNN-1 compared to the spectroscopic ground truth for the GalaxiesML test sample of $\sim28,000$ galaxies; in addition, the predictions by NN-2 and BNN-2 are compared to the photometric ground truth for the TransferZ sample of $\sim11,000$ galaxies. Qualitatively, the deterministic NN (NN-1 and NN-2) show better performance than their probabilistic counterparts. The BNNs, in particular, demonstrate a predictive degradation after $z\sim2$ where training data are scarce in both GalaxiesML and TransferZ. We report the baseline benchmark performance of deterministic and probabilistic NNs in Table \ref{tab:metric_fnn} and \ref{tab:metric_bnn} under the heading "Precision". These metrics are evaluated within the $0.3<z<1.5$ redshift range. NN-1 demonstrates 6.1 times lower bias ($\nnBiasGMLone$ versus $\BnnBiasGMLone \times10^{-3}$), 1.3 times lower outlier rate ($\nnOGMLone\%$ versus $\BnnOGMLone\%$), and 1.3 times less scatter than BNN-1 ($\nnScatGMLone$ versus $\BnnScatGMLone \times10^{-2}$). Similarly, NN-2 outperforms BNN-2 with 4.8 times less bias ($\nnBiasTZtwo$ versus $\BnnBiasTZtwo \times10^{-3})$, 1.6 times lower outlier rate ($\nnOTZtwo\%$ versus $\BnnOTZtwo\%$), and 1.3 times less scatter ($\nnScatTZtwo$ versus $\BnnScatTZtwo\times10^{-2}$). However, the probabilistic NNs (BNN-1 and BNN-2) produce 3 times lower $3\sigma$ outlier rate than the deterministic NN equivalent models ($\BnnOsigConfGMLone\%$ versus $\nnOsigConfGMLone\%$ and $\BnnOsigConfTZtwo\%$ versus $\nnOsigConfTZtwo\%$). Since NNs do not inherently provide uncertainty estimates, split conformal prediction generated prediction intervals enable measurement of the $3\sigma$ outlier rate and a direct comparison to the probabilistic BNN models.

Most models achieve a precision that meets the LSST science requirement for bias and $3\sigma$ outlier rate within the $0.3 < z < 1.5$ redshift range when evaluated on their respective training ground truths. Figure \ref{fig:nn_metric_comparison} and \ref{fig:bnn_metric_comparison} show the performance metrics for the deterministic and probabilistic baseline models, respectively. Values within the gray shaded area satisfy the LSST science requirements. Single ground-truth trained models, with the exception of BNN-2, meet the bias and $3\sigma$ outlier rate LSST science requirements on their respective samples, but fail to meet the scatter requirement. BNN-2 only satisfies the $3\sigma$ outlier rate requirement.

\subsection{Photo-\textit{z} Generalization: Training and Evaluating on Different Data}

\begin{figure*}
    \centering
    \includegraphics[width=\textwidth]{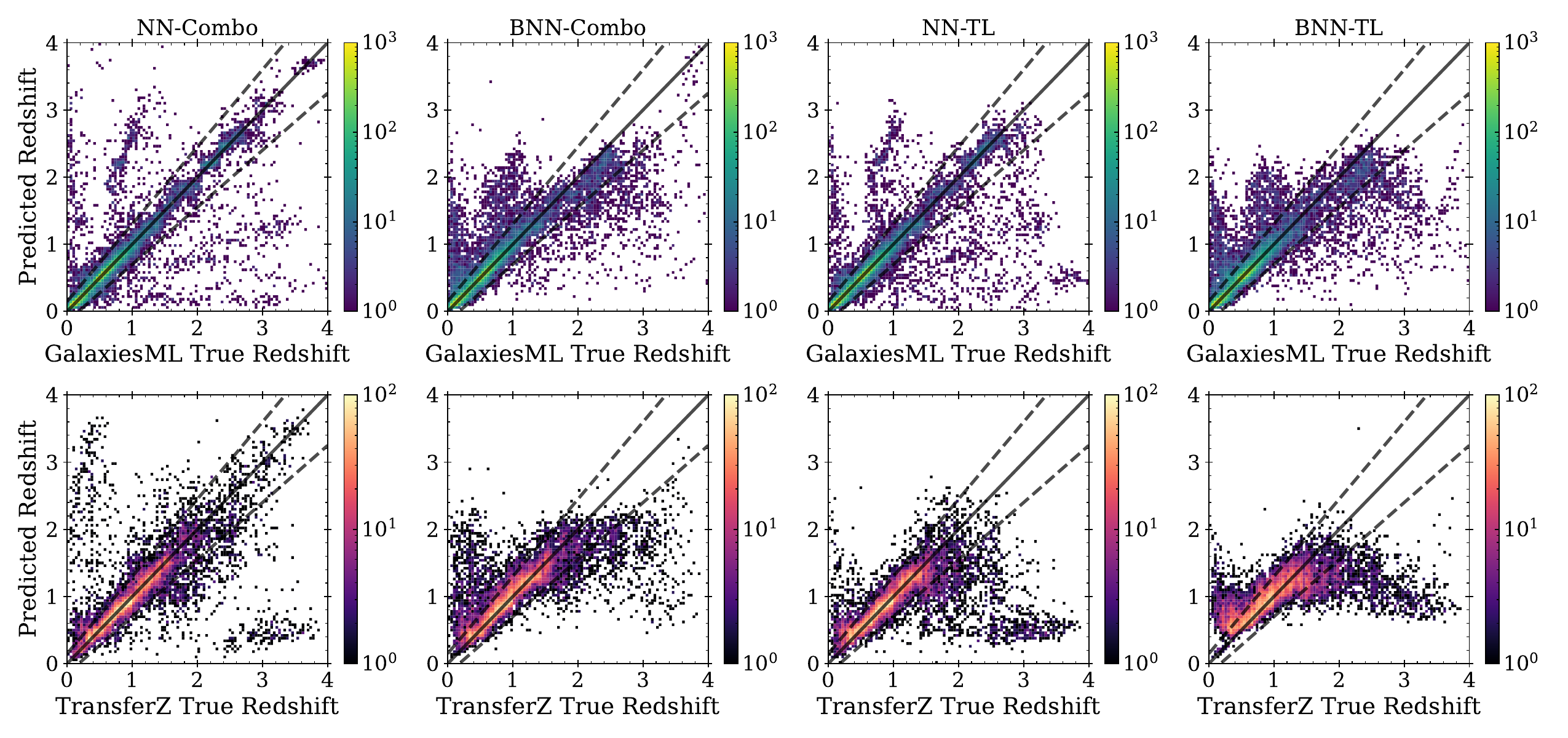}
    \caption{Photometric vs. ground-truth redshift for deterministic and probabilistic models trained on different approaches to mixing of GalaxiesML and TransferZ. The models are evaluated on a 28,000 sample of galaxies from GalaxiesML (top) and 11,000 sample of galaxies from TransferZ (bottom). The NN-Combo and BNN-Combo are models trained on a composite dataset approach while NN-TL and BNN-TL are trained on transfer learning from a base model trained on TransferZ and fine-tuned on GalaxiesML. The solid black line is a one-to-one line and the dashed black lines correspond to outliers with $z_{\rm pred}>\pm0.15(1+z_{\rm true})$. The color scale indicates the density of data points (log scale). The deterministic models show less scatter and lower bias on both datasets than probabilistic models, with BNN-TL showing the most scatter.}
    \label{fig:mixed_pred}
\end{figure*}

Models trained on GalaxiesML show substantial performance degradation when evaluated on TransferZ in comparison to evaluation with GalaxiesML. The first two panels in the bottom row of Figure \ref{fig:single_pred} show the predictions of NN-1 and BNN-1 on the TransferZ test sample of $\sim11,000$ galaxies. Within the $0.3<z<1.5$ redshift range, the deterministic NN model evaluated on TransferZ results in 127 times more bias ($\nnBiasTZone$ versus $\nnBiasGMLone\times10^{-3}$), 2.9 times higher outlier rate ($\nnOTZone\%$ versus $\nnOGMLone\%$), and 2.5 times more scatter ($\nnScatTZone$ versus $\nnScatGMLone\times10^{-2}$) than on GalaxiesML. Similarly, the probabilistic NN model has similar degradation, with 28 times more bias ($\BnnBiasTZone$ versus $\BnnBiasGMLone\times10^{-3}$), 2.7 times more scatter ($\BnnScatTZone$ versus $\BnnScatGMLone\times10^{-2}$), and 2.5 times higher outlier rate ($\BnnOTZone\%$ versus $\BnnOGMLone\%$) when comparing TransferZ evaluation against GalaxiesML. When restricting the TransferZ sample to brighter galaxies ($i<20.5$), the spectroscopic ground-truth trained model predictions have comparable scatter and outlier rates as when evaluated on the GalaxiesML sample, but $\sim$25 times higher bias. We expect this degradation in performance given that NN-1 and BNN-1 are trained on the limited galaxy population in GalaxiesML, which is not representative of the more diverse galaxy types found in TransferZ.

When evaluated on GalaxiesML, a model trained on TransferZ shows comparable results to its evaluation on TransferZ. These models do not perform better than the baseline trained on GalaxiesML. The last two panels in the bottom row of Figure \ref{fig:single_pred} show the predictions of NN-2 and BNN-2 on the GalaxiesML test sample of $\sim$28,000 galaxies. For both NN-2 and BNN-2, most predictions lie within the outlier bounds when $z_{\rm truth}<1.5$. Turning to a quantitative assessment, for the $0.3<z<1.5$ redshift range, we compare prediction metrics on GalaxiesML to the results on TransferZ (Table \ref{tab:metric_fnn} and \ref{tab:metric_bnn}). NN-2 shows mixed results, with 1.3 times less scatter ($\nnScatGMLtwo$ versus $\nnScatTZtwo\times10^{-2}$) on GalaxiesML compared to TransferZ, but 1.6 times higher bias ($\nnBiasGMLtwo$ versus $\nnBiasTZtwo\times10^{-3}$) and 1.2 times higher outlier rate ($\nnOGMLtwo\%$ versus $\nnOTZtwo\%$). BNN-2 performs more consistently, with 2.1 times less bias ($\BnnBiasGMLtwo$ versus $\BnnBiasTZtwo\times10^{-3}$), 1.6 times less scatter ($\BnnScatGMLtwo$ versus $\BnnScatTZtwo\times10^{-2}$), and 1.4 times lower outlier rate ($\BnnOGMLtwo\%$ versus $\BnnOTZtwo\%$). The photometric ground-truth trained models learn the bright galaxy population for the redshift range $0.3<z_{\rm truth}<1.5$ from TransferZ alone. Both models display poor predictions within $0.4<z_{\rm truth}<4$ and $z_{\rm pred}<0.2$ caused by quasars present in GalaxiesML but absent from their TransferZ training set, leading to misclassifications of these sources as low-redshift galaxies.  

Evaluating models on a ground truth different from that seen during training fails to meet LSST science requirements (Fig. \ref{fig:nn_metric_comparison} and \ref{fig:bnn_metric_comparison}). For spectroscopic ground-truth models evaluated on TransferZ, BNN-1 satisfies a single requirement (the $3\sigma$ outlier rate), but no other requirement. NN-1 does not meet any of the requirements. For photometric ground-truth models evaluated on GalaxiesML, the results depend on model architecture. The deterministic model (NN-2) satisfies the $3\sigma$ outlier rate requirement, but the probabilistic model does not meet the requirement. Both NN-2 and BNN-2 fail to meet the bias and scatter requirements. Overall, these models fail to generalize in the context of LSST science requirements.

\subsection{Approaches to Combining Ground-Truth Performance}
So far, we have looked at single ground-truth trained models evaluated on spectroscopic and photometric samples. We now turn our focus to the two approaches of combining ground truths: direct combination or transfer learning. For the direct-combination approach, the NN-Combo and BNN-Combo models are trained on 281,000 galaxies from a mixture of GalaxiesML and TransferZ. For the transfer-learning approach, the NN-TL and BNN-TL models are fine-tuned on 200,000 GalaxiesML galaxies, using NN-2 and BNN-2 as baselines for the deterministic and probabilistic models, respectively. Figure \ref{fig:mixed_pred} shows the model predictions on the GalaxiesML test sample of 28,000 galaxies (top row) and on TransferZ test sample of 11,000 galaxies (bottom row). The predictions reveal how each approach handles the ground-truth combination challenge. 

The composite dataset approach performance is dependent on the model architecture for both ground truths. In Figure \ref{fig:mixed_pred}, NN-Combo maintains consistent predictions across the full redshift range for both GalaxiesML and TransferZ test samples, while BNN-Combo shows good performance only within $0.3<z_{\rm truth}<1.5$ before degrading substantially at higher redshifts. In the redshift range $0.3<z<1.5$, the NN-Combo scatter is $\sigma_{\rm NMAD}=\nnScatGMLcom\times10^{-2}$ on GalaxiesML, which is 1.9 times less than $\sigma_{\rm NMAD}=\nnScatTZcom\times10^{-2}$ on TransferZ, with an outlier rate 1.7 times lower on GalaxiesML than TransferZ ($\nnOGMLcom\%$ versus $\nnOTZcom\%$). Also, the NN-Combo bias of $|b|=\nnBiasGMLcom\times10^{-3}$ on GalaxiesML is 3.2 times lower than $|b|=\nnBiasTZcom\times10^{-3}$ on TransferZ. On the other hand, BNN-Combo performs significantly worse on TransferZ than GalaxiesML, with a 251 times higher higher bias ($\BnnBiasTZcom$ versus $\BnnBiasGMLcom\times10^{-3})$, 2.1 times more scatter ($\BnnScatTZcom$ versus $\BnnScatGMLcom\times10^{-2}$), and 1.8 times higher outlier rate ($\BnnOTZcom\%$ versus $\BnnOGMLcom\%$). Overall, the deterministic architecture trained on the composite dataset is more robust on both ground truths than its probabilistic counterpart.

Our transfer-learning approach learns the spectroscopic ground truth with greater accuracy across metrics than the photometric ground truth. The NN-TL and BNN-TL predictions on TransferZ are highly scattered with a degradation at high redshifts ($z_{\rm truth}>1.5$); on the other hand, the models predict GalaxiesML redshifts well with low scatter and stronger predictions up to $z_{\rm truth}>2$ (Fig. \ref{fig:mixed_pred}). Quantitatively, the deterministic and probabilistic model bias on GalaxiesML is 5.3 times less ($\nnBiasGMLtl$ versus $\nnBiasTZtl\times10^{-3}$) and 17.9 times less ($\BnnBiasGMLtl$ versus $\BnnBiasTZtl\times10^{-3}$) than evaluations on TransferZ, respectively; furthermore, the models show 2 times less scatter on GalaxiesML than TransferZ ($\nnScatGMLtl$ versus $\nnScatTZtl\times10^{-2}$ and $\BnnScatGMLtl$ versus $\BnnScatTZtl\times10^{-2}$, respectively). Lastly, NN-TL and BNN-TL outlier rates on GalaxiesML ($\nnOGMLtl\%$ and $\BnnOGMLtl\%$) are 1.4 times and 2.3 times lower than on TransferZ ($\nnOTZtl\%$ and $\BnnOTZtl\%$), respectively. This asymmetric performance indicates models trained on the transfer-learning approach favor spectroscopic ground truth over the photometric ground truth. However, the deterministic model demonstrates more balanced performance than the probabilistic version on both ground truths.  

All models meet the LSST bias and outlier science requirements for the evaluation on GalaxiesML within the redshift range $0.3<z<1.5$; however, when evaluated on TransferZ, the success in meeting the criteria is architecture dependent (Fig. \ref{fig:nn_metric_comparison} and \ref{fig:bnn_metric_comparison}). On GalaxiesML, the science requirements are met with bias values $0.07-1.96$ ($\times10^{-3}$), all below the threshold of $3\times10^{-3}$, and $3\sigma$ outlier rates less than 10\%, ranging from $2.8\%$ to $8.2\%$. However, the scatter values $2.4-3.5$ ($\times10^{-2}$) are greater than the requirement of $2\times10^{-2}$, with NN-Combo producing the lowest scatter. With TransferZ, the architecture strengths determine which accuracy metrics are met by the models. The probabilistic models BNN-TL and BNN-Combo meet the $3\sigma$ outlier rate requirements $\sim2.5\%$ ($<10\%$ requirement), but the deterministic model counterparts do not meet this requirement. For the bias requirement on TransferZ, NN-Combo meets the requirement but the other mixed ground-truth models do not meet this requirement. All models fail to meet the scatter requirement.

\subsection{Uncertainty Quantification}

In the previous section, we reported the $3\sigma$ outlier rate results for both deterministic and probabilistic models after applying split conformal prediction. In this section, we validate the predicted photo-\textit{z} uncertainty estimates calibrated with conformal prediction for the BNNs and prediction intervals generated for the NNs. We focus on the metrics detailed in Section \ref{sec:prob_metrics}. In particular, we focus on uncertainty estimates for each ground truth. 

\begin{figure}
    \centering
    \includegraphics[width=\linewidth]{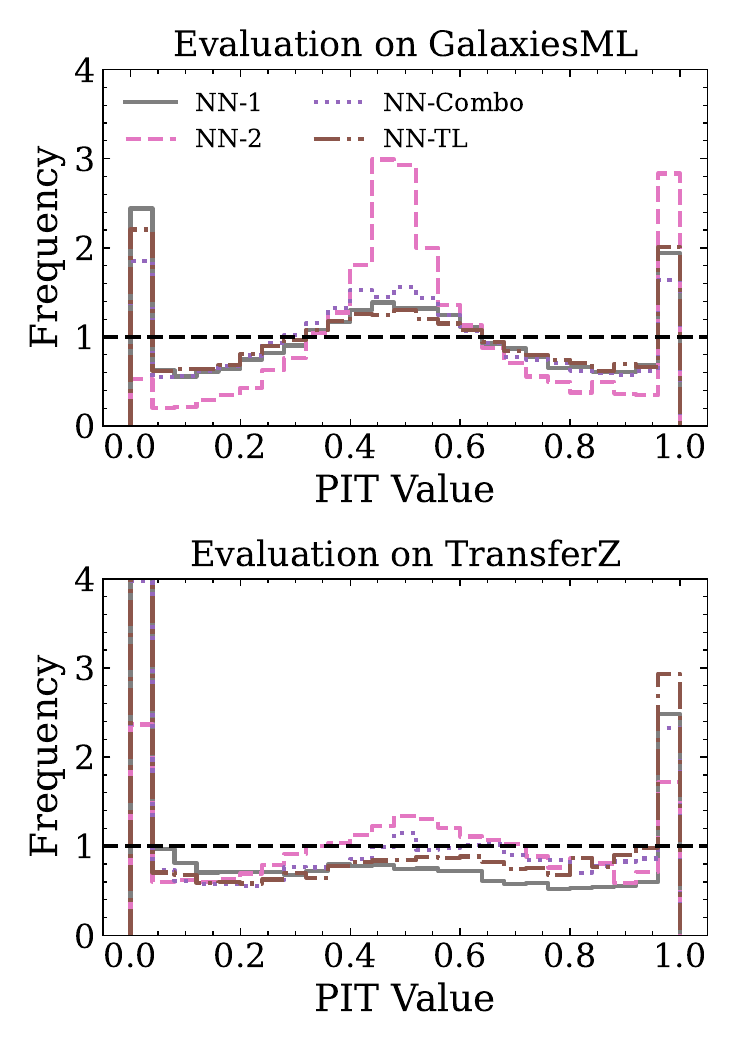}
    \caption{Probabilistic calibration assessment of GalaxiesML (top) and TransferZ (bottom) predicted pdfs by deterministic NNs after creating prediction intervals using split conformal prediction. The probability integral transform (PIT) is a qualitative assessment of pdfs, with good calibration indicated by a uniform distribution of \textit{U}[0,1]. The GalaxiesML PIT distributions are center-peaked, indicating overdispersed pdfs, while the TransferZ PIT distributions are U-shaped, indicating underdispersed PDFs. The presence of peaks at the edges indicates the presence of outliers.}
    \label{fig:nn_pit}
\end{figure}

Prediction intervals created by split conformal prediction for deterministic models produce miscalibrated pdfs. Figure \ref{fig:nn_pit} shows the PIT distributions for the GalaxiesML and TransferZ samples using the deterministic models. For GalaxiesML, all traditional NN models display a distribution with a peak at the center and edges, indicating overdispersed pdfs and outliers. The NN-2 distribution is the most prominent distribution with the highest peaks at the center. The PIT distribution for the TransferZ sample shows the inverse, with a U-shaped distribution, indicating underdispersed pdfs. This behavior is expected because the prediction intervals are assigned using the median of the residuals within a $\Delta z=0.1$ redshift bin. 

\begin{figure}
    \centering
    \includegraphics[width=\linewidth]{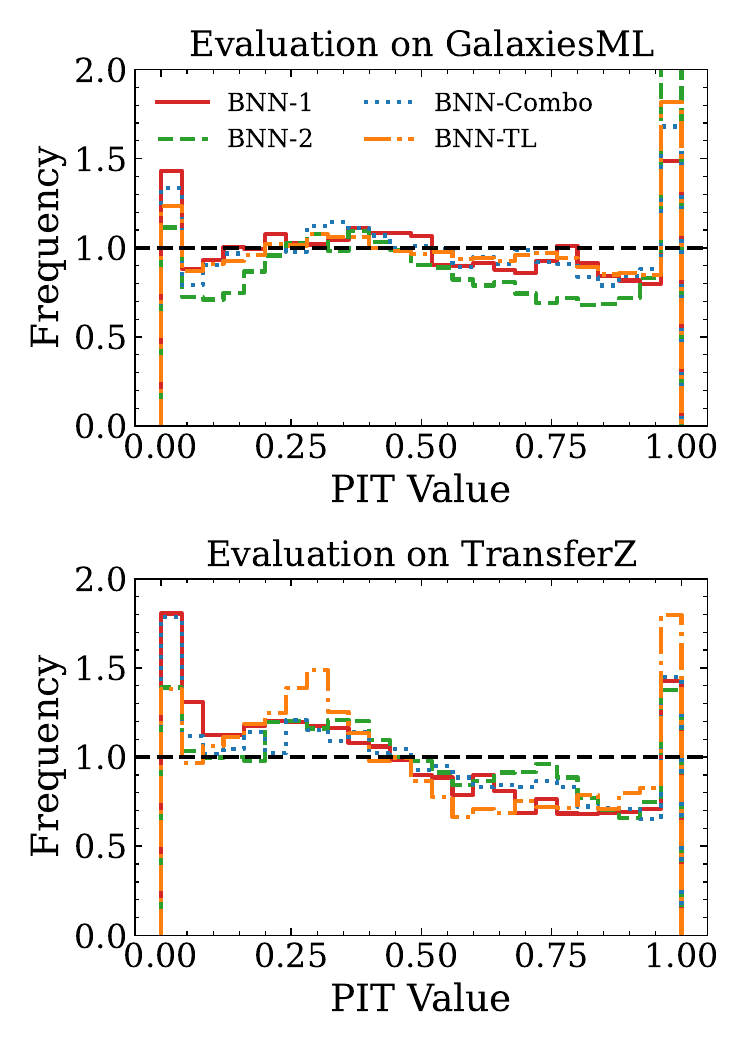}
    \caption{Probabilistic calibration assessment of GalaxiesML (top) and TransferZ (bottom) predicted pdfs by BNNs after applying split conformal prediction. The probability integral transform (PIT) is a qualitative assessment of pdfs with good calibration indicated by a uniform distribution \textit{U}[0,1]. Most GalaxiesML PIT distributions are uniform with BNN-2 showing a U-shaped distribution, an indication of underdispersed pdfs. On the other hand, the TransferZ PIT distributions are right-skewed, indicating the presence of a systematic bias. All PIT distributions have spikes at the edges caused by outliers; the height of these spikes is model and data dependent.}
    \label{fig:bnn_pit}
\end{figure}

Our probabilistic NNs trained on GalaxiesML (only or a mixture) produce reliable pdfs while the pdfs for TransferZ are poor. Figure \ref{fig:bnn_pit} shows the PIT distributions for GalaxiesML and TransferZ for the pdfs measured by the BNNs, where the black dash line indicates a uniform (ideal) PIT distribution. For GalaxiesML, the BNN models with an exception of BNN-2 hover around the uniform \textit{U}[0,1] line, indicating accurate predictions with reliable pdfs. The BNN-2 model displays a U-shaped distribution, indicating a model overconfidence, which is expected as the BNN-2 is not trained on GalaxiesML. The higher peak at PIT$\,=1$ than PIT$\,= 0$ is caused by both the underestimation seen in the previous section and this overconfidence. The peaks at PIT$\,=1$ are present for other BNNs, but are significantly smaller. For TransferZ, all BNN PIT distributions decrease from low to high PIT values, indicating a systematic bias in the pdfs. This feature is caused by overestimation of the true redshift, overconfident uncertainty estimates, or both; decoupling the cause is nontrivial. Thus, the PIT distributions reveal that the BNNs (except for BNN-2) produce reliable pdfs for spectroscopic sources, though all BNNs are miscalibrated for TransferZ. The uncertainty estimates are most reliable for BNN models trained only on GalaxiesML or using a mixed ground-truth approach.

\begin{figure}
    \centering
    \includegraphics[width=\linewidth]{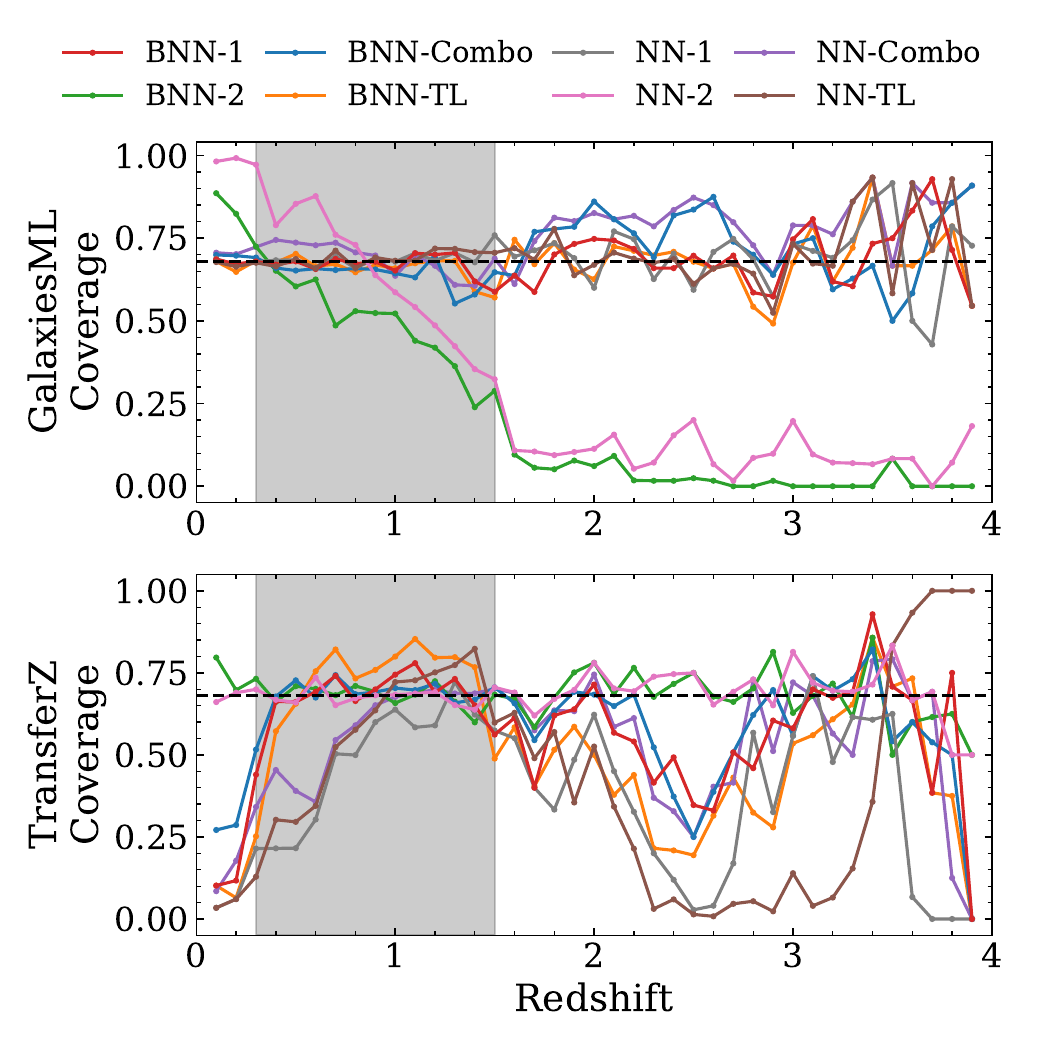}
    \caption{Statistical coverage –- the fraction of galaxies that have their ground-truth redshift within their 68\% prediction interval after split conformal prediction --- plots for all models.  Within the redshift range $0.3 < z < 1.5$, BNN-Combo and BNN-1 maintain the most consistent statistical coverage across both datasets. Models trained on a single dataset tend to achieve statistical coverage but struggle to calibrate on a different dataset.}
    \label{fig:coverage_plot}
\end{figure}

Our models, with the exception of those trained only on photometric redshifts, achieve the statistical coverage level of $68\%$ on GalaxiesML within the redshift range $0.3<z<1.5$. However, only BNNs and NN-2 achieve the target coverage level on TransferZ. Figure \ref{fig:coverage_plot} shows the statistical coverage for the GalaxiesML and TransferZ samples using all the models across the full range of our samples. Models maintain a stable coverage level for GalaxiesML at low redshifts $0<z<1.5$. However, their statistical coverage deviates more at higher redshifts. This trend is expected as fewer training data are present at these redshifts, so uncertainty estimates will be less stable at higher redshifts. The BNN-2 and NN-2 models display statistical coverage below the $68\%$ coverage level, indicating, similar to the PIT distribution, overconfident or underdispersed uncertainty estimates. On the other hand, the NN-2 and BNN-2 statistical coverage is stable for the TransferZ sample for the full redshift range. Similarly, the BNNs achieve the required statistical coverage levels, with instabilities at $z<0.3$ and $z>1.5$. The deterministic models, NN-1, NN-TL, and NN-Combo, show statistical coverage below the $68\%$ level, indicating underdispersed uncertainty estimates. The statistical coverage indicates reliable uncertainty estimates produced by BNNs for both ground truths. 

\begin{figure}
    \centering
    \includegraphics[width=\columnwidth]{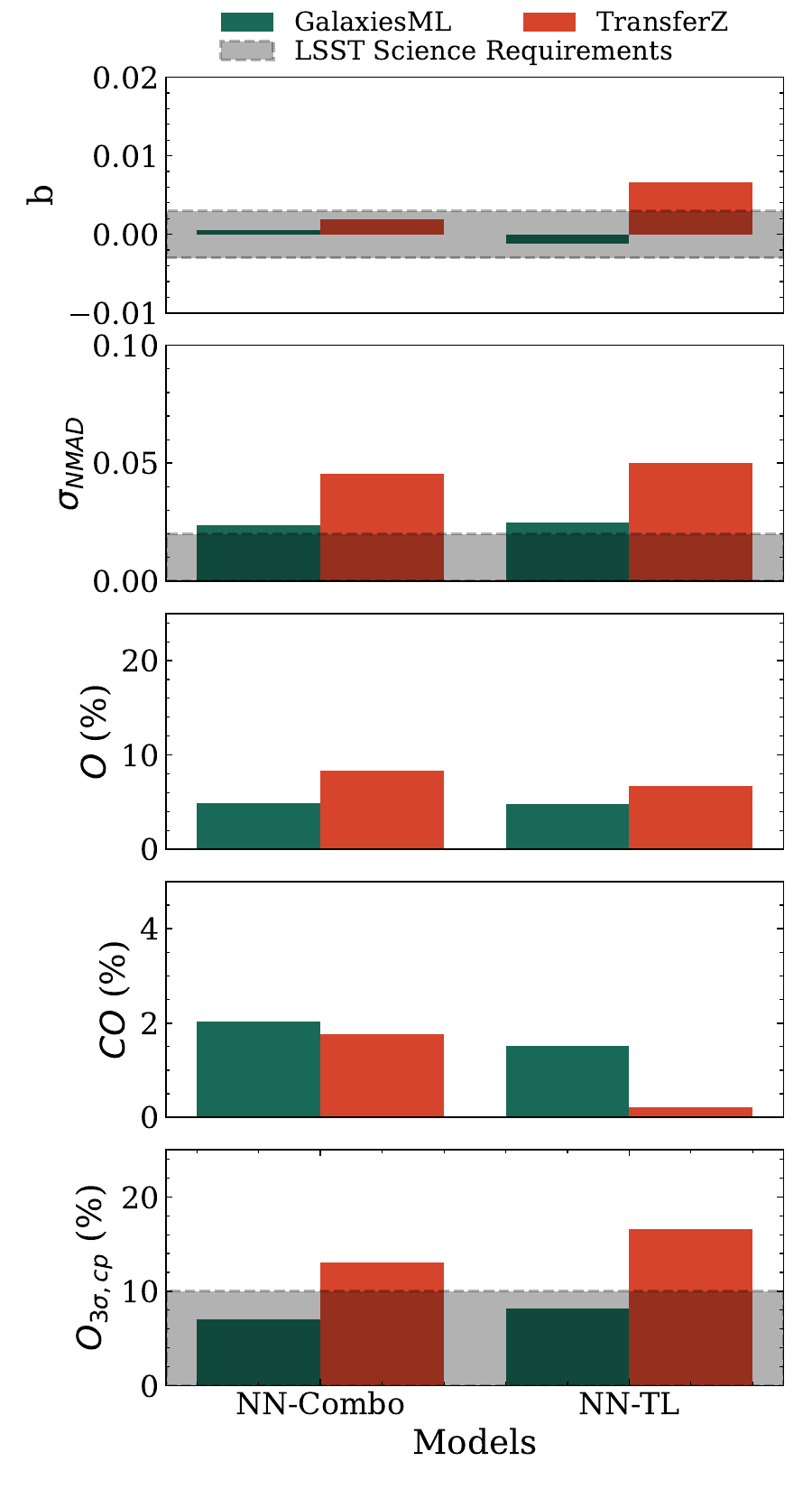}
    \caption{A comparison of the performance between two of our approaches for combining ground truths using metrics from Sections \ref{sec:point_metrics} and \ref{sec:prob_metrics}. We evaluate the performance of redshift predictions averaged over the redshift range $0.3<z<1.5$. The two chosen approaches are transfer learning (NN-TL) and composite dataset training (NN-Combo). NN-Combo performs better than NN-TL across both datasets on LSST-like metrics.}.
    \label{fig:tl_versus_combo}
\end{figure}

Overall, we find the best model depends on the use case. The best photo-\textit{z} performance on both ground truths is achieved by the deterministic model trained on a composite dataset (NN-Combo). On both ground-truth samples, NN-Combo shows the most consistent performance in comparison to our other models (Fig. \ref{fig:tl_versus_combo}). However, NN-Combo does not produce the most reliable uncertainty estimates. The most reliable uncertainty estimates are produced by the BNNs (with the exception of BNN-2) on both ground truths.

\section{Discussion}
In this work, we evaluate the effectiveness of two approaches to combining different sources of ground truth for photometric redshift estimation: transfer learning and composite dataset training. We find that training a deterministic neural network on a composite dataset (NN-Combo) is better than both transfer learning and models trained on single ground-truth datasets. This work generalizes our findings from \citet{soriano2024} to additional photo-\textit{z} metrics and the case of a BNN; in addition, this work systematizes a comparison of probabilistic neural networks to the performance of neural networks trained on just GalaxiesML and TransferZ individually. We note that \citet{soriano2024} found the transfer learning to be effective, however this was with regard to a model trained only on TransferZ. 

The NN-Combo model performs at a similar level to other photo-\textit{z} estimation techniques. We note that a comparison with these techniques is not necessarily straightforward, as the setups are not identical and are typically not evaluated over similar redshift ranges or with similar goals. \citet{tanaka2018} compared different photo-\textit{z} codes using a training sample similar to GalaxiesML, and reported a scatter of $\sigma_{\text{NMAD}}=0.05$ and an outlier rate of $15\%$ for galaxies brighter than $i<25$ and between $0.3<z<1.5$. Our NN-Combo approach achieves $\sigma_{\text{NMAD}} = 0.024$ with a $\nnOGMLcom\%$ outlier rate in a comparable magnitude and redshift range, demonstrating better performance under similar evaluation conditions. A complementary approach by \citet{cabayol2023} used multitask learning to improve broadband photometric redshifts, achieving a $\sigma=0.0163$ and outlier fraction of 0.5\% up to $z<1.5$. Their study, which utilized COSMOS2015 \citep{laigle2016} photo-\textit{z} estimates as training labels to extend their spectroscopic sample, demonstrates another effective strategy for enhancing redshift estimation performance. While their methodology differs from our composite dataset approach, both studies highlight the benefits of incorporating diverse photometric information. 

Our work shows that NNs outperform BNNs in maintaining consistent results across the different ground truths. BNNs provide the significant advantage of producing uncertainty estimates that are reliable and well calibrated after split conformal prediction. However, BNN's complex architecture appears to make them sensitive to distributional differences between spectroscopic and photometric redshifts. These differences are most pronounced on transfer-learning models, whose performance is an order of magnitude worse on the BNN than the NN. In contrast, the simpler architecture of NN models offer more consistent performance; however when supplemented with split conformal prediction for uncertainty quantification, they produce overdispersed pdfs. These architectural distinctions are important to cosmological probes, as accurate characterization of the photo-\textit{z} pdfs is equally critical for cosmological studies as the performance \citep{newman2022}. Recent work on incorporating galaxy images directly into NN architectures has shown promising improvements in photometric redshift estimation and uncertainty quantification \citep{jones2024a}. Such approaches could potentially reduce the distribution mismatch we observed and enhance model performance across diverse galaxy populations. 

There are a few other limitations that need to be addressed in future work. Figure \ref{fig:single_pred} shows a distinct feature in the predictions for the NN-2 and BNN-2 models, which we find corresponds to a population of quasars not present in TransferZ. We determined this classification using a positional crossmatch of these sources to the existing literature. Thus, future work could seek to update TransferZ with this population, potentially improving mixed ground-truth trained models. In addition, the transfer learning approach showed greater accuracy on the spectroscopic ground truth over the photometric ground truth. Because these models are fine-tuned on the spectroscopic ground truth from GalaxiesML, the models' knowledge of the photometric ground truth from TransferZ appears to diminish. We explore only a subset of the full hyperparameter space for these transfer learning models, so future work could look into different implementations of this technique.

Furthermore, the implicit normality assumption used when characterizing photo-\textit{z} uncertainties poses a limitation. For LSST-like data, photometric redshift pdfs are frequently non-Gaussian. Most prevalent for faint sources, these pdfs often display long tails, skewness, or multimodality. This simplification may influence downstream metrics that depend on the pdf shape, including interpretation of PIT distributions and calibration behavior of split conformal prediction. Future work could relax this assumption by adopting a flexible pdf for the BNNs.

Since the TransferZ ground truth uses photometric redshifts as ground-truth  labels, the reported errors likely underestimate the true errors. Through error propagation these errors can be a factor $1.5-2$ times higher for predictions on the TransferZ test set. Despite this limitation, TransferZ provides valuable insights into model generalization across different galaxy populations, and demonstrates the utility of photometric redshift labels for extending training samples to underrepresented parameter space.

\section{Conclusion}
With next-generation large-scale surveys coming on-line, development of photo-\textit{z} estimation models will enable advancements in weak-lensing and dark energy studies. These models must balance performance (precision of redshift estimates) and characterization (recovery of redshift distributions via uncertainty quantification) over diverse galaxy populations. Traditional machine learning models calibrated on highly precise spectroscopic redshift samples that are often biased toward bright galaxies and limited in color-color space coverage. To generalize to a broader sample of galaxies, models must incorporate reliable photometric redshifts that probe wider color-color space and fainter populations, with evaluation frameworks aligned with survey requirements like those of LSST.

In this work, we explored the challenge of generalizing photometric redshift estimation across different ground-truth sources using neural networks combined with split conformal prediction. We created TransferZ, which includes photometric redshifts derived from the COSMSO2020 catalog and photometry from the HSC-SSP survey, as a complementary dataset to the spectroscopic redshifts in GalaxiesML. We evaluated two approaches to combining ground truths --- transfer learning and training on a composite dataset --- against single ground-truth trained neural networks. Our results demonstrate that a neural network trained on a composite dataset (NN-Combo) is the most consistent across datasets. It meets two out of the three LSST science requirements within the redshift range $0.3<z<1.5$ when evaluated on GalaxiesML, and satisfies one out of three requirements for TransferZ. NN-Combo shows promising generalization across models trained on different ground truths, performing well on both samples, though not yet at LSST-required levels. As the first data release for LSST approaches, photo-\textit{z} models must perform well over a broad population of galaxies for precision cosmology. We hope the framework of this paper is helpful for improving photo-\textit{z} prediction and uncertainties by leveraging sources of ground truth that can complement spectroscopic redshifts. 

\section{acknowledgments}
Partial support for this work was provided by the Alfred P. Sloan Foundation and the NSF DGE2034835. This work used Jetstream2 at Indiana University through allocation No. PHY230092 from the Advanced Cyberinfrastructure Coordination Ecosystem: Services $\&$ Support (ACCESS) program, which is supported by U.S. National Science Foundation grant Nos. 2138259, 2138286, 2138307, 2137603, and 2138296. 

The Hyper Suprime-Cam (HSC) collaboration includes the astronomical communities of Japan and Taiwan, and Princeton University. The HSC instrumentation and software were developed by the National Astronomical Observatory of Japan (NAOJ), the Kavli Institute for the Physics and Mathematics of the Universe (Kavli IPMU), the University of Tokyo, the High Energy Accelerator Research Organization (KEK), the Academia Sinica Institute for Astronomy and Astrophysics in Taiwan (ASIAA), and Princeton University. Funding was contributed by the FIRST program from the Japanese Cabinet Office, the Ministry of Education, Culture, Sports, Science and Technology (MEXT), the Japan Society for the Promotion of Science (JSPS), Japan Science and Technology Agency (JST), the Toray Science Foundation, NAOJ, Kavli IPMU, KEK, ASIAA, and Princeton University. 

This paper makes use of software developed for the Large Synoptic Survey Telescope. We thank the LSST Project for making their code available as free software at  http://pipelines.lsst.io/.

This paper is based (in part) on data collected at the Subaru Telescope and retrieved from the HSC data archive system, which is operated by the Subaru Telescope and Astronomy Data Center (ADC) at National Astronomical Observatory of Japan. Data analysis was in part carried out with the cooperation of Center for Computational Astrophysics (CfCA), National Astronomical Observatory of Japan. The Subaru Telescope is honored and grateful for the opportunity of observing the Universe from Maunakea, which has the cultural, historical and natural significance in Hawaii. 

This work is based on observations collected at the European Southern Observatory under
ESO program ID 179.A-2005 and on data products produced by CALET and the Cambridge Astronomy Survey Unit on behalf of the UltraVISTA consortium.    

\bibliography{project.bib}{}
\bibliographystyle{aasjournal}

\appendix
\section{Defining Reliable Photometric Redshift Measurements}\label{sec:appendix_a}
To develop machine learning models for photometric redshift estimation, we required a training sample with reliable redshift measurements. We define reliable photometric redshifts similar to \citet{singal2022}, who applied their criteria to COSMOS2015 \citep{laigle2016}. These criteria were designed to ensure a resulting dataset that is optimal for machine learning applications. Our criteria for reliable photometric redshifts are as follows:
\begin{enumerate}
    \item Sources must have photometric redshift estimates present in both catalog versions (\texttt{Classic} and \texttt{The Farmer}) and for both codes (\texttt{LePhare} and \texttt{EAZY}).
    \item Sources are required to have magnitude measurements between 0 and 50 in both catalog versions for 22 of the 35 bands.
    \item Sources needed a reported $\chi^2<1$ and $\chi^2<2.08$ for the \texttt{LePhare} estimates in \texttt{Classic} and \texttt{The Farmer}, respectively. In addition, the peak of the pdf and the photometric redshift that minimizes the $\chi^2$ must be comparable, $|\texttt{zPDF} - \texttt{zMinChi2}|<0.1$.
    \item All photometric redshifts must fall within $z\leq4$.
    \item All other photometric redshift estimate sets must agree with the \texttt{Classic} catalog's \texttt{LePhare} estimates within a difference of $0.1$.
\end{enumerate}

We note that the four catalog/photo-\textit{z} code combinations utilize up to 25-35 filters in their measurements. Therefore, criterion 2 is based on identifying the common photometric coverage (25 bands) across all catalog/code combinations. We also account for the Lyman break drop in flux, so criterion 2 excludes the CFHT \textit{U} bands and bluest Suprime-Cam narrow band.

\section{Neural Network Architectures}\label{sec:appendix_b}
\begin{figure}[h]
    \centering
    \includegraphics[width=0.7\linewidth]{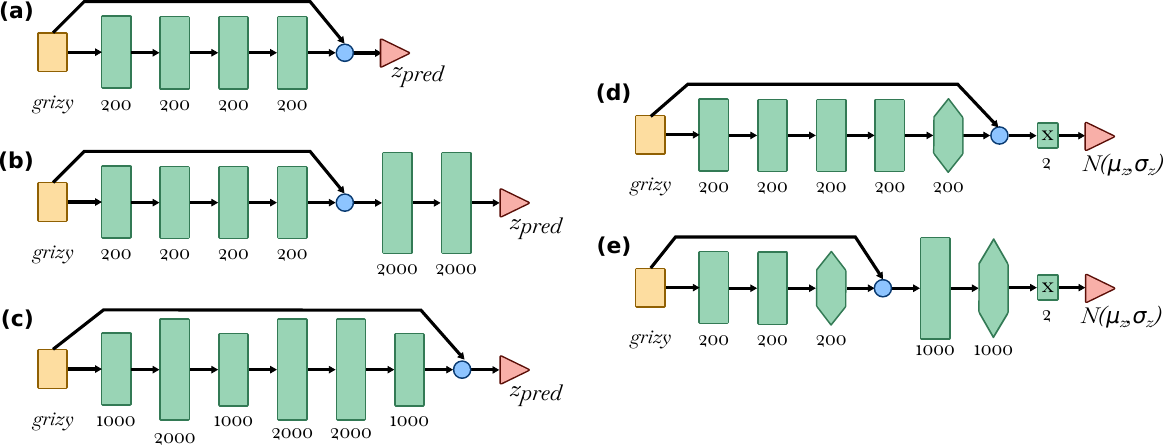}
    \caption{Neural network (NN) model architectures used throughout this work. Rectangles correspond to regular dense layers with their corresponding number of nodes, elongated hexagons correspond to probabilistic dense variational layers, and circles correspond to skip connections. All hidden layers use ReLU activation functions except for those with an "x," indicating no activation function is utilized. All models have five-band \textit{grizy} magnitudes as inputs and output redshift predictions. We give more detail on the implementation in Section \ref{sec:architectures}. Architecture (a) is for NN-1, (b) for NN-2 and NN-TL, (c) for NN-Combo, (d) for BNN-1 and (e) for BNN-2, BNN-TL, and BNN-Combo.}
    \label{fig:model_architectures}
\end{figure}

Here, we present a visualization of the architectures used to train the eight models (Fig. \ref{fig:model_architectures}). For more detail on the implementation, see Section \ref{sec:architectures}.

\section{Uncertainty estimates before and after split conformal prediction}\label{sec:appendix_c}

\begin{figure}[h]
    \centering
    \includegraphics[width=0.6\linewidth]{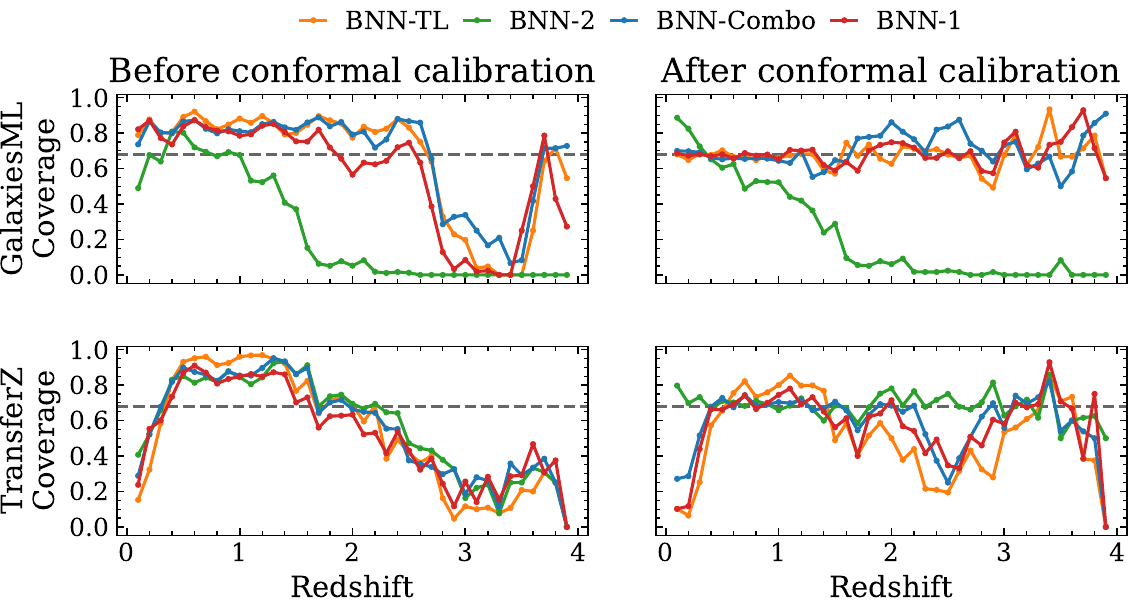}
    \caption{Comparison of statistical coverage per tomographic bin before and after applying split conformal prediction to BNN predictions. Ideally, $68\%$ of evaluated galaxies will have the ground-truth redshift within their $68\%$ prediction interval. A percentage of galaxies above this target level indicates galaxies are "overcovered" or have wide prediction intervals. A percentage of galaxies below this target level indicate "undercovered" or narrow prediction intervals. Before split conformal prediction, BNN evaluated galaxies from the GalaxiesML and TransferZ samples produce wide prediction intervals at low redshift and tend to become narrow at higher redshifts. After split conformal prediction, the desired statistical coverage level is achieved at the full redshift range for GalaxiesML; however, only BNN-2 achieves accurate coverage on the TransferZ sample.}
    \label{fig:coverage_compare}
\end{figure}

The full, accurate redshift distribution of a sample is necessary in tomographic weak-lensing analyses \citep{ma2006}. BNNs produce pdfs for individual objects, but these pdfs can be poorly calibrated. Figure \ref{fig:coverage_compare} shows the $1\sigma$ coverage in tomographic bins for $0<z<4$ spaced by $\Delta z=0.1$ before and after applying split conformal prediction using all BNNs trained in this work. For each tomographic bin, the statistical coverage is not accurate on either dataset. The uncertainty estimates are too wide for $z<2$ on GalaxiesML and $0.3<z<1.6$, with a trend toward narrow uncertainty estimates for higher redshifts. The BNN-2 model is trained on the TransferZ sample, leading to poor statistical coverage on the GalaxiesML sample; however, the statistical coverage on TransferZ follows a similar pattern to the other models. After split conformal prediction, the models show accurate statistical coverage throughout the redshift range on the GalaxiesML sample. Similar to before split conformal prediction, the BNN-2 model statistical coverage is poor on GalaxiesML, but it is accurate on TransferZ. The statistical coverage of the other BNNs is improved in the redshift range $0.3<z<1.5$, but narrow uncertainty estimates remain for other tomographic bins. Thus, applying split conformal prediction has a positive impact in ensuring reliable pdfs where a substantial representative sample is available (see Fig. \ref{fig:dataset_comparison}).

\begin{figure}[h]
    \centering
    \includegraphics[width=0.6\linewidth]{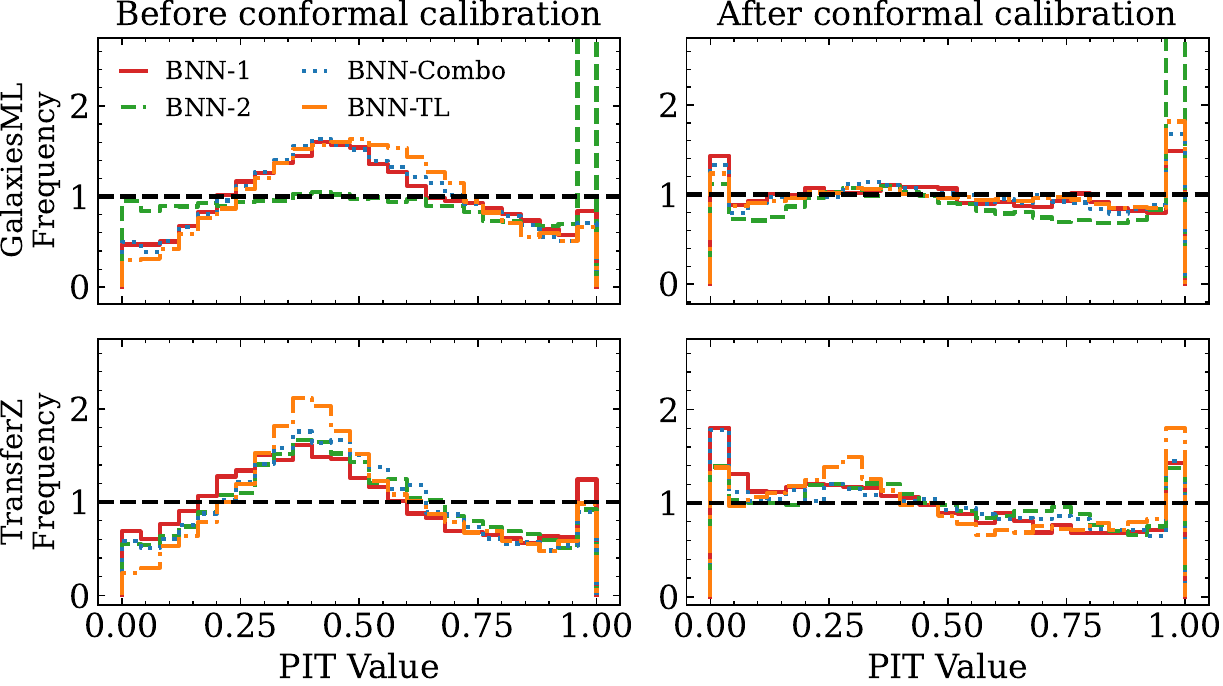}
    \caption{Comparison of the PIT distribution before and after applying split conformal prediction on evaluated galaxies from the GalaxiesML and TransferZ samples. The evaluations are performed using the BNN models developed in this work. The PIT distribution is a qualitative diagnostic of accurate pdfs over an ensemble of galaxies. Accurate pdfs are indicated by a uniform PIT distribution \textit{U}[0,1]. Our application of split conformal prediction ensures the pdfs are reliable. Before split conformal prediction, the PIT distributions are center-peaked, an indication of overdispersed pdfs; after split conformal prediction, the distributions are closer to uniform.}
    \label{fig:pit_compare}
\end{figure}

Figure \ref{fig:pit_compare} shows the PIT distributions for the GalaxiesML and TransferZ samples before and after split conformal prediction using our BNN models. In accordance with the statistical coverage plot, the center-peaked PIT distributions before split conformal prediction indicate overdispersed pdfs; after split conformal prediction, the distributions are more uniform for GalaxiesML, but with some indications of systematic issues with pdfs predicted for the TransferZ sample.  

\section{Binned Photo-\textit{z} performance}\label{sec:metric_per}

\begin{figure}[h]
    \centering
    \includegraphics[width=0.8\linewidth]{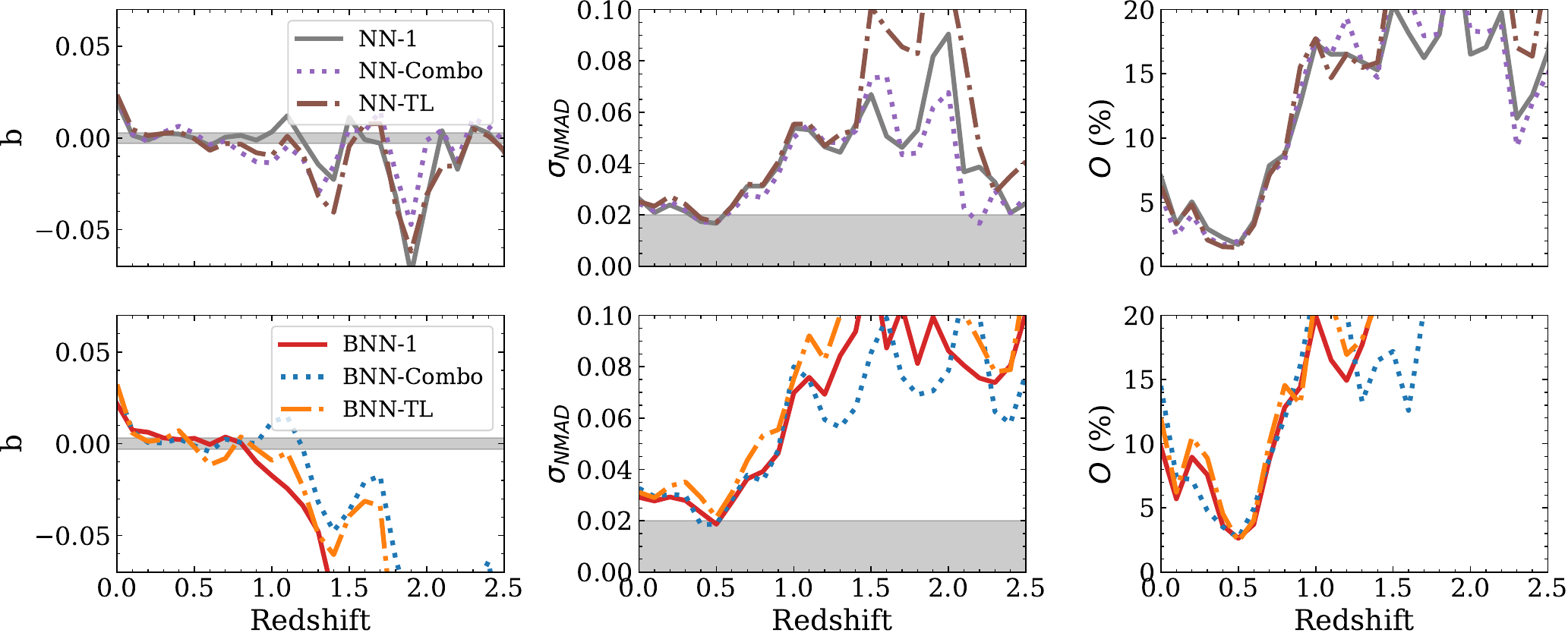}
    \caption{Photo-\textit{z} performance on bias (left column), scatter (middle column), and outlier rate (right column) for equally spaced redshift bins ($\delta z=0.01$) up to $z<2.5$ for the NNs (top row) and BNNs (bottom row) evaluated on the GalaxiesML test set. The gray shaded areas correspond to the LSST science goals. Regardless of the training methodology, the models perform similarly across this redshift range.}
    \label{fig:metric_per_z}
\end{figure}

\begin{figure}[h]
    \centering
    \includegraphics[width=0.8\linewidth]{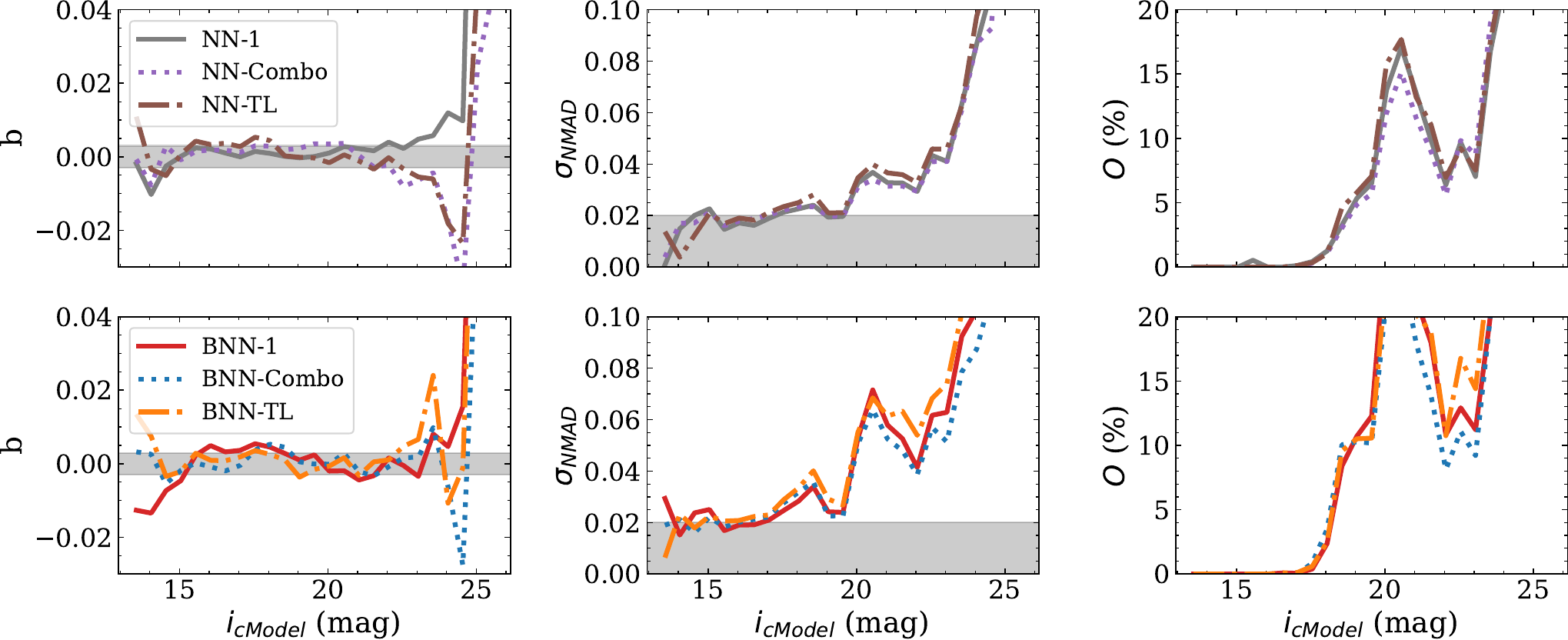}
    \caption{Photo-\textit{z} performance on bias (left column), scatter (middle column), and outlier rate (right column) for equally spaced magnitude bins ($\delta i=0.5$) up to $i<25$ mag for the NNs (top row) and BNNs (bottom row) evaluated on the GalaxiesML test set. The gray shaded areas correspond to the LSST science goals. The bias performance for all models is around the LSST bias requirement for sources brighter than $i<23$ mag. The photo-\textit{z} scatter and outlier rates increase consistently for fainter sources.}
    \label{fig:metric_per_mag}
\end{figure}

We present the photometric redshift performance metrics as a function of redshift and magnitude. Figure \ref{fig:metric_per_z} and \ref{fig:metric_per_mag} show the photo-\textit{z} performance (bias, scatter, and outlier rate) of our models evaluated on the GalaxiesML test set with $z<2.5$ and $i<25$ mag. We exclude NN-2 and BNN-2 from this analysis.

Standard NN models follow similar trends, as do all BNN models; however, the two model architectures show distinct behavior with respect to the bias metric. The NN models show oscillatory patterns across redshift, maintaining $|b|<0.02$ for $z<1.2$ but often exceeding this limit for $z>1.2$. These oscillations likely reflect the transition of spectral features (e.g. the $4000$\AA\, break) through the HSC photometric bands. In contrast, the BNN models show less oscillatory bias behavior, maintaining $|b|<0.02$ for $z<1.2$. Nevertheless, all BNNs exhibit consistently negative bias (systematic underprediction) at higher redshifts. As a function of magnitude, the bias remains well controlled with $|b|<0.015$ for $i<23$ mag
across all models (NNs and BNNs). At fainter magnitudes, all model performance degrades.
 
All models show similar trends in scatter and outlier rate, with both metrics increasing at high redshifts and faint magnitudes. For low-redshift sources ($z<0.5$) and bright sources ($i<20$ mag), all models achieve $\sigma_{\rm NMAD}<0.04$ and maintain outlier fractions below $O<10\%$. Scatter increases severely at $z>1.5$ and $i>24$ mag for NNs and at $z>1.0$ and $i>23$ mag for BNNs, where training data become scarce. However, the outlier behavior in redshift space is model dependent. BNN-TL and BNN-Combo maintain $O<10\%$ in a narrow range $0.3<z<0.6$, struggling with low-redshift galaxies, while NN models and BNN-1 maintain this threshold for $z<0.7$.

\section{Model Predictions}\label{sec:mrt}

Here, we present the photometric redshift predictions from all eight neural network models developed in this work (Table \ref{tab:predictions}). For each of the 40,269 galaxies in the combined GalaxiesML and TransferZ test samples, we provide the HSC-PDR2 \textit{grizy} magnitudes (columns 3-7), ground truth redshifts (column 8), and predicted redshifts with associated uncertainties from both the deterministic NNs and BNNs (columns 9-25). 

\movetabledown=2.5in
\begin{rotatetable}
\begin{deluxetable}{rrrrrrrrrrrrrrrrrrrrrrrrr}
\tabletypesize{\tiny}
\tablecolumns{25}
\tablecaption{Example photometric redshift estimates from our neural network.}
\tablewidth{0pt}
\setlength{\tabcolsep}{0.5mm}
\tablehead{
\colhead{ID1} & \colhead{ID2} & \colhead{\textit{g}} & \colhead{\textit{r}} & \colhead{\textit{i}} & \colhead{\textit{z}} & \colhead{\textit{y}} & \colhead{$z_{\rm truth}$} & \colhead{Source} & \colhead{$z_{\rm NN1}$} & \colhead{$\sigma_{\rm z,NN1}$}  & \colhead{$z_{\rm NN2}$} & \colhead{$\sigma_{\rm z,NN2}$}  & \colhead{$z_{\rm NNTL}$} & \colhead{$\sigma_{\rm z,NNTL}$}  & \colhead{$z_{\rm NNC}$} & \colhead{$\sigma_{\rm z,NNC}$}  & \colhead{$z_{\rm BNN1}$} & \colhead{$\sigma_{\rm z,BNN1}$}  & \colhead{$z_{\rm BNN2}$} & \colhead{$\sigma_{\rm z,BNN2}$}  & \colhead{$z_{\rm BNN-TL}$} & \colhead{$\sigma_{\rm z,BNN-TL}$}  & \colhead{$z_{\rm BNNC}$} & \colhead{$\sigma_{\rm z,BNNC}$} \\
\colhead{(1)} & \colhead{(2)} & \colhead{(3)} & \colhead{(4)} & \colhead{(5)} & \colhead{(6)} & \colhead{(7)} & \colhead{(8)} &\colhead{(9)} & \colhead{(10)} & \colhead{(11)} & \colhead{(12)} & \colhead{(13)} & \colhead{(14)} & \colhead{(15)} & \colhead{(16)} & \colhead{(17)} & \colhead{(18)} & \colhead{(19)} & \colhead{(20)} & \colhead{(21)} & \colhead{(22)} & \colhead{(23)} & \colhead{(24)} & \colhead{(25)}
}
\startdata
43158876522051997 &    1109991 & 23.314 & 22.405 & 21.579 & 21.320 & 21.067 &  0.699 &       1 & 0.698 &   0.05 & 0.698 &   0.07 & 0.703 &    0.05 &    0.683 &       0.05 & 0.720 &    0.05 & 0.679 &    0.04 &  0.688 &     0.06 &     0.699 &        0.05 \\
42692142425968305 &         -1 & 21.646 & 20.737 & 19.974 & 19.627 & 19.390 &  0.658 &       0 & 0.643 &   0.05 & 0.625 &   0.07 & 0.639 &    0.05 &    0.637 &       0.05 & 0.649 &    0.09 & 0.628 &    0.04 &  0.633 &     0.13 &     0.636 &        0.10 \\
43158322471270138 &     929890 & 24.462 & 23.988 & 23.849 & 23.457 & 23.932 &  0.185 &       1 & 0.812 &   0.03 & 0.303 &   0.28 & 0.511 &    0.03 &    0.349 &       0.03 & 0.727 &    0.41 & 0.545 &    0.86 &  0.745 &     0.35 &     0.704 &        0.40 \\
43347567320240912 &         -1 & 19.675 & 18.993 & 18.647 & 18.455 & 18.325 &  0.051 &       0 & 0.174 &   0.04 & 0.152 &   0.55 & 0.156 &    0.04 &    0.168 &       0.04 & 0.218 &    0.19 & 0.176 &    0.14 &  0.159 &     0.11 &     0.165 &        0.08 \\
43153658136785393 &    1337086 & 23.884 & 23.336 & 22.582 & 22.270 & 22.273 &  0.847 &       1 & 0.816 &   0.07 & 0.830 &   0.07 & 0.825 &    0.07 &    0.803 &       0.06 & 0.824 &    0.05 & 0.815 &    0.04 &  0.817 &     0.08 &     0.813 &        0.03 \\
70404207722662055 &         -1 & 19.836 & 18.431 & 17.904 & 17.604 & 17.409 &  0.273 &       0 & 0.255 &   0.04 & 0.263 &   0.19 & 0.262 &    0.04 &    0.262 &       0.05 & 0.273 &    0.02 & 0.292 &    0.08 &  0.260 &     0.02 &     0.257 &        0.02 \\
43159155694895835 &    1164719 & 23.640 & 22.673 & 22.302 & 22.213 & 22.047 &  0.462 &       1 & 0.524 &   0.03 & 0.471 &   0.06 & 0.523 &    0.03 &    0.491 &       0.04 & 0.534 &    0.11 & 0.428 &    0.06 &  0.574 &     0.12 &     0.462 &        0.05 \\
43158464205165381 &    1034964 & 25.960 & 24.179 & 22.958 & 22.126 & 21.809 &  1.018 &       1 & 0.963 &   0.14 & 0.972 &   0.12 & 0.959 &    0.15 &    0.977 &       0.13 & 0.967 &    0.06 & 0.986 &    0.05 &  0.961 &     0.08 &     0.989 &        0.06 \\
42630428040920903 &         -1 & 18.655 & 17.759 & 17.289 & 16.911 & 16.687 &  0.149 &       0 & 0.107 &   0.03 & 0.158 &   0.28 & 0.111 &    0.03 &    0.111 &       0.03 & 0.102 &    0.02 & 0.159 &    0.07 &  0.120 &     0.02 &     0.113 &        0.03 \\
41241203689155979 &         -1 & 18.208 & 17.376 & 16.854 & 16.520 & 16.259 &  0.044 &       0 & 0.090 &   0.04 & 0.153 &   0.55 & 0.093 &    0.04 &    0.090 &       0.04 & 0.090 &    0.02 & 0.124 &    0.11 &  0.103 &     0.03 &     0.096 &        0.03 \\
\enddata
\tablecomments{Column (1): HSC-PDR2 Wide identifier. Column (2): COSMOS2020 Classic Catalog identifier. '-1' corresponds to sources not in the catalog. Column (3): data source for galaxy, '0' for the data from GalaxiesML and '1' for the data from TransferZ. Columns (4)-(8): HSC-PDR2 \textit{grizy} magnitudes. Column (9) true redshift of galaxy. Columns (10)-(11): predicted redshift and conformal prediction interval, respectively, for galaxy using NN-1. Columns (12)-(13): predicted redshift and conformal prediction interval, respectively, for galaxy using NN-2. Columns (14)-(15): predicted redshift and conformal prediction interval, respectively, for galaxy using NN-TL. Columns (16)-(17): predicted redshift and conformal prediction interval, respectively, for galaxy using NN-Combo. Columns (18)-(19): predicted redshift and conformal prediction interval, respectively, for galaxy using BNN-1. Columns (20)-(21): predicted redshift and conformal prediction interval, respectively, for galaxy using BNN-2. Columns (22)-(23): predicted redshift and conformal prediction interval, respectively, for galaxy using BNN-TL. Columns (24)-(25): predicted redshift and conformal prediction interval, respectively, for galaxy using BNN-Combo.
(This table is available in its entirety in a machine-readable form.)}
\end{deluxetable}   
\end{rotatetable}
\label{tab:predictions}
\clearpage

\end{document}

%% file: metrics.tex
\def\nnBiasGMLone{0.17}
\def\nnScatGMLone{2.3}
\def\nnOGMLone{5.1}
\def\nnCOGMLone{1.6}
\def\nnOsigConfGMLone{9.0}

\def\nnBiasTZone{21.71} 
\def\nnScatTZone{5.7}
\def\nnOTZone{14.8}
\def\nnCOTZone{2.0}
\def\nnOsigConfTZone{23.0}

\def\nnBiasCone{3.97} 
\def\nnScatCone{3.2}
\def\nnOCone{8.4}
\def\nnCOCone{1.7}
\def\nnOsigConfCone{13.1}

\def\nnBiasGMLtwo{4.56}
\def\nnScatGMLtwo{3.4}
\def\nnOGMLtwo{9.6}
\def\nnCOGMLtwo{2.1}
\def\nnOsigConfGMLtwo{9.3}

\def\nnBiasTZtwo{2.88} 
\def\nnScatTZtwo{4.3}
\def\nnOTZtwo{7.7}
\def\nnCOTZtwo{1.5}
\def\nnOsigConfTZtwo{7.3}

\def\nnBiasCtwo{2.41} 
\def\nnScatCtwo{3.7}
\def\nnOCtwo{9.0}
\def\nnCOCtwo{1.9}
\def\nnOsigConfCtwo{8.7}

\def\nnBiasGMLtl{1.24}
\def\nnScatGMLtl{2.5}
\def\nnOGMLtl{4.8}
\def\nnCOGMLtl{1.5}
\def\nnOsigConfGMLtl{8.2}

\def\nnBiasTZtl{6.60} 
\def\nnScatTZtl{5.0}
\def\nnOTZtl{6.7}
\def\nnCOTZtl{0.2}
\def\nnOsigConfTZtl{16.6}

\def\nnBiasCtl{0.24} 
\def\nnScatCtl{3.2}
\def\nnOCtl{5.5}
\def\nnCOCtl{1.1}
\def\nnOsigConfCtl{10.6}

\def\nnBiasGMLcom{0.59}
\def\nnScatGMLcom{2.4}
\def\nnOGMLcom{4.9}
\def\nnCOGMLcom{2.0}
\def\nnOsigConfGMLcom{7.0}

\def\nnBiasTZcom{1.86} 
\def\nnScatTZcom{4.5}
\def\nnOTZcom{8.3}
\def\nnCOTZcom{1.8}
\def\nnOsigConfTZcom{13.1}

\def\nnBiasCcom{0.89} 
\def\nnScatCcom{2.9}
\def\nnOCcom{6.1}
\def\nnCOCcom{1.9}
\def\nnOsigConfCcom{8.7}

\def\BnnBiasGMLone{1.04}
\def\BnnScatGMLone{2.9}
\def\BnnOGMLone{6.8}
\def\BnnCOGMLone{0.8}
\def\BnnOsigConfGMLone{2.7}

\def\BnnBiasTZone{29.23} 
\def\BnnScatTZone{7.7}
\def\BnnOTZone{17.0}
\def\BnnCOTZone{1.6}
\def\BnnOsigConfTZone{3.1}

\def\BnnBiasCone{5.53} 
\def\BnnScatCone{4.1}
\def\BnnOCone{10.3}
\def\BnnCOCone{1.1}
\def\BnnOsigConfCone{2.8}

\def\BnnBiasGMLtwo{6.43}
\def\BnnScatGMLtwo{3.6}
\def\BnnOGMLtwo{9.2}
\def\BnnCOGMLtwo{2.0}
\def\BnnOsigConfGMLtwo{12.7}

\def\BnnBiasTZtwo{13.80} 
\def\BnnScatTZtwo{5.8}
\def\BnnOTZtwo{12.7}
\def\BnnCOTZtwo{1.7}
\def\BnnOsigConfTZtwo{2.3}

\def\BnnBiasCtwo{1.41} 
\def\BnnScatCtwo{4.1}
\def\BnnOCtwo{10.4}
\def\BnnCOCtwo{1.9}
\def\BnnOsigConfCtwo{9.7}

\def\BnnBiasGMLtl{1.96}
\def\BnnScatGMLtl{3.5}
\def\BnnOGMLtl{7.4}
\def\BnnCOGMLtl{0.9}
\def\BnnOsigConfGMLtl{2.8}

\def\BnnBiasTZtl{35.17} 
\def\BnnScatTZtl{8.2}
\def\BnnOTZtl{16.8}
\def\BnnCOTZtl{0.1}
\def\BnnOsigConfTZtl{3.0}

\def\BnnBiasCtl{4.83} 
\def\BnnScatCtl{4.8}
\def\BnnOCtl{10.7}
\def\BnnCOCtl{0.6}
\def\BnnOsigConfCtl{2.9}

\def\BnnBiasGMLcom{0.07}
\def\BnnScatGMLcom{2.8}
\def\BnnOGMLcom{6.8}
\def\BnnCOGMLcom{1.2}
\def\BnnOsigConfGMLcom{3.1}

\def\BnnBiasTZcom{17.60} 
\def\BnnScatTZcom{5.9}
\def\BnnOTZcom{12.5}
\def\BnnCOTZcom{1.6}
\def\BnnOsigConfTZcom{3.0}

\def\BnnBiasCcom{3.09} 
\def\BnnScatCcom{3.6}
\def\BnnOCcom{8.7}
\def\BnnCOCcom{1.3}
\def\BnnOsigConfCcom{3.1}